\documentstyle[11pt,aaspp4, graphics]{article}    

\slugcomment{To appear in the Astrophysical Journal}
\lefthead{Lockitch, Friedman}
\righthead{}

\newcommand{\half}{{\case{1}{2}}}
\newcommand{\be}{\begin{equation}}
\newcommand{\ee}{\end{equation}}
\newcommand{\ba}{\begin{eqnarray}}
\newcommand{\ea}{\end{eqnarray}}
\newtheorem{thm}{Theorem}

\begin{document}


\title{Where are the r-modes of isentropic stars?}
\author{Keith H. Lockitch and John L. Friedman}
\affil{University of Wisconsin-Milwaukee, P.O. Box 413, Milwaukee, WI 53201 
 \\ lockitch@uwm.edu, friedman@uwm.edu}
\begin{abstract}
Almost none of the r-modes ordinarily found in rotating stars exist, if
the star and its perturbations obey the same one-parameter equation of
state; and rotating relativistic stars with one-parameter equations of
state have no pure r-modes at all, no modes whose limit, for a star
with zero angular velocity, is a perturbation with axial parity. Similarly
(as we show here) rotating stars of this kind have no pure g-modes, no modes
whose spherical limit is a perturbation with polar parity and vanishing
perturbed pressure and density. Where have these modes gone?

In spherical stars of this kind, r-modes and g-modes form a degenerate
zero-frequency subspace.  We find that rotation splits the degeneracy
to {\it zeroth} order in the star's angular velocity $\Omega$, and the
resulting modes are generically hybrids, whose limit as
$\Omega\rightarrow 0$ is a stationary current with axial and  polar
parts.  Lindblom and Ipser have recently found these hybrid modes in an
analytic study of the Maclaurin spheroids.  Since the hybrid modes have 
a rotational restoring force, they call them ``rotation modes'' or 
``generalized r-modes''.

Because each mode has definite parity, its axial and polar
parts have alternating values of $l$.   We show that each mode belongs
to one of two classes, axial-led or polar-led, depending on
whether the spherical harmonic with lowest value of $l$ that
contributes to its velocity field is axial or polar. We numerically
compute these modes for slowly rotating polytropes and for Maclaurin
spheroids, using a straightforward method that appears to be novel and
robust. Timescales for the gravitational-wave driven instability and
for viscous damping are computed using assumptions appropriate to
neutron stars.  The instability to nonaxisymmetric modes is, as
expected, dominated by the $l=m$ r-modes with simplest radial
dependence, the only modes which retain their axial character in 
isentropic models;  for relativistic isentropic stars, these $l=m$
modes must also be replaced by hybrids of the kind considered here.
\end{abstract}



\section{Introduction}

A recently discovered instability of r-modes of rotating stars (first
found in numerical study by Andersson (1998), and analytically verified
by Friedman and Morsink (1998)) has gained the attention of a number of
authors (\cite{k98}; Lindblom, Owen, and Morsink 1998; Owen et al. 1998; 
Andersson, Kokkotas and Schutz 1998; \cite{ks98}; \cite{akst98}; Lindblom 
and Ipser 1998; Madsen 1998; \cite{s99}; \cite{bk99}; \cite{lmo99}).
These modes have axial parity (see below) and their frequency is 
proportional to the star's angular velocity. Neutron stars that are rapidly 
rotating at birth are likely to be unstable to nonaxisymmetric perturbations 
driven by gravitational waves; estimates of growth times and viscous damping 
times (Lindblom et al 1998, \cite{o98}, Andersson et al 1998, \cite{ks98}, 
\cite{lmo99}) suggest that r-modes dominate the spin-down of such stars for 
several months, until a superfluid transition shuts off the instability.  
Unstable r-modes may thus set the upper limit on the spin of young neutron 
stars, and gravitational waves emitted during the initial spin-down might 
be detectable.  The recent discovery by Marshall et al (1998) of a pulsar 
in the supernova remnant N157B implies the existence of a class of neutron 
stars that are rapidly rotating at birth and whose spin is plausibly limited 
by the gravitational-wave driven instability.

Perturbations of a spherical star can be divided into two classes,
axial and polar, depending on their behavior under parity.  Where polar
tensor fields on a 2-sphere can be constructed from the scalars $Y_l^m$
and their gradients $\nabla Y_l^m$ (and the metric on a 2-sphere),
axial fields involve the pseudo-vector $\hat r\times \nabla Y_l^m$, and
their behavior under parity is opposite to that of $Y_l^m$.  That is,
axial perturbations of odd $l$ are invariant under parity, and axial
perturbations with even $l$ change sign. If a mode varies continuously
along a sequence of equilibrium configurations that starts with a
spherical star and continues along a path of increasing rotation, the
mode will be called axial if it is axial for the spherical star.  Its
parity cannot change along the sequence, but $l$ is well-defined only
for modes of the spherical configuration.

It is useful to further divide stellar perturbations into subclasses
according to the physics dominating their behaviour. In perfect
fluid stellar models, the polar parity perturbations consist of the 
f-, p- and g-modes; the f- and p-modes having pressure as their dominant
restoring force and the g-modes having gravity as their dominant restoring 
force (Cowling 1941).  The axial parity perturbations are dominated by the
Coriolis force in rotating stars and were called ``r-modes'' by 
Papaloizou and Pringle (1978) because of their similarity to the Rossby 
waves of terrestrial meteorology.  For slowly rotating stars, the r-modes
have frequencies which scale linearly with the star's angular velocity and 
in the spherical limit they become time-independent convective currents.

Despite the sudden interest in these modes, however, they are not
yet well-understood for stellar models in which both the star and 
its perturbations are governed by a one-parameter equation of state, 
$p=p(\rho)$; we shall call such stellar models isentropic, because 
isentropic models and their adiabatic perturbations obey the same 
one-parameter equation of state.  For stars with more general equations of 
state, the r-modes appear to be complete for perturbations that have axial 
parity.  However, this is not the case for isentropic models.  One finds 
that the only purely axial modes allowed in isentropic stars are the 
physically interesting $l=m$ r-modes with simplest radial behavior 
(\cite{pp78};  Provost et al. 1978\footnote{An appendix in this paper 
incorrectly claims that no $l=m$ r-modes exist, based on an incorrect 
assumption about their radial behavior}; \cite{s82}; \cite{sm83}). 
The disappearance of the purely axial modes with $l>m$ occurs for the 
following reason.  In spherical isentropic stars the gravitational 
restoring forces that give rise to the g-modes vanish and they, too,
become time-independent convective currents with vanishing perturbed
pressure and density.  Thus, the space of zero frequency modes, which 
generally consists only of the axial r-modes, becomes larger for 
spherical isentropic stars to include the polar g-modes.  This large 
degenerate subspace of zero-frequency modes is split by rotation to zeroth 
order in the angular velocity, and the corresponding modes of rotating 
isentropic stars are hybrids whose spherical limits are mixtures of axial 
and polar perturbations.  These hybrid modes have already been found 
analytically for the uniform-density Maclaurin spheroids by Lindblom and 
Ipser (1998) who point out that since their dominant restoring force is 
the Coriolis force, it is natural to refer to them as rotation modes, or 
generalized r-modes.

Although isentropic Newtonian stars do retain a vestigial set of purely axial 
modes (those having $l=m$), it appears that rotating relativistic stars of 
this type have {\it no} pure r-modes, no modes whose limit for a spherical 
star is purely axial (Andersson, Lockitch and Friedman, 1999).
For nonisentropic relativistic stars, Kojima (1998) has derived an equation
governing purely axial perturbations to lowest order in the star's angular 
velocity.\footnote{Based on this equation, Kojima has argued that the 
spectrum is continuous, and his argument has been made precise in a recent 
paper of Beyer and Kokkotas (1999) (See also Kojima and Hosonuma 1999).  
Beyer and Kokkotas, however, also point 
out that the continuous spectrum they find may be an artifact of the fact that 
the imaginary part of the frequency vanishes in the slow-rotation limit.}
For the isentropic case, however, we find a second, independent equation for 
these perturbations that appears to give inconsistent radial behaviour for
purely axial modes (Andersson, Lockitch and Friedman, 1999). This second 
equation is obtained from an angular component of the curl of the 
relativistic Euler equation and is simply the relativistic generalization
of the equations presented in Sect. III of this paper 
(Eq. (\ref{qphi2}), for example).  Kojima's equation is one from which 
polar-parity perturbations have been excluded, and, as the present paper 
makes clear, one cannot assume that axial and polar parity modes decouple 
for slowly rotating isentropic stars. Instead, we expect the Newtonian $l=m$ 
r-modes to become discrete axial-led hybrids of the corresponding 
relativistic models.  

In this paper we examine the hybrid rotational modes of rotating isentropic 
Newtonian stars.  We distinguish two types of modes, axial-led and
polar-led, and show that every mode belongs to one of the two classes.  
We then turn to the computation of eigenfunctions and eigenfrequencies for 
modes in each class, adopting what appears to be a method that is both novel and
robust.  For the uniform-density Maclaurin spheroids, these modes have 
been found analytically by Lindblom and Ipser in a complementary 
presentation that makes certain features transparent but masks properties 
that are our primary concern.  We examine the eigenfrequencies and 
corresponding eigenfunctions to lowest nontrivial order in the angular 
velocity $\Omega$.  We then examine the frequencies
and modes of $n=1$ polytropes, finding that the structure of the modes
and their frequencies are very similar for the polytropes and the
uniform-density configurations.  The numerical analysis is complicated by
a curious linear dependence in the Euler equations, detailed in Appendix B.
The linear dependence appears in a power series expansion of the equations 
about the origin.  It may be related to difficulty other groups have 
encountered in searching for these modes.

Finally, we examine unstable modes, computing their growth time and 
expected viscous damping time.  The pure $l=m=2$ r-mode retains its 
dominant role, but the $3\leq l=m\lesssim10$ r-modes and some of the 
fastest growing hybrids may contribute to the gravitational radiation 
and spin-down. 


\section{Spherical Stars}

We consider a static spherically symmetric, self-gravitating perfect 
fluid described by a gravitational potential $\Phi$, density $\rho$ and 
pressure $p$. These satisfy an equation of state of the form
\be
p = p( \rho), \label{eos}
\ee 
as well as the Newtonian equilibrium equations
\be
\nabla_a(h + \Phi) = 0  \label{equil}
\ee
\be
\nabla^2 \Phi = 4\pi G\rho, \label{newton} \label{equil_end}
\ee
where $h$ is the specific enthalpy in a comoving frame,
\be
h = \int \frac{dp}{\rho}.
\ee 

We are interested in the space of zero-frequency modes, the linearized 
time-independent perturbations of this static equilibrium.  This 
zero-frequency subspace is spanned by two types of perturbations: (i)
perturbations with $\delta v^a\neq 0$ and $\delta \rho = \delta p = 
\delta \Phi = 0$, and (ii) perturbations with $\delta \rho$, $\delta p$ and 
$\delta \Phi$ nonzero and $\delta v^a=0$.  If one assumes that no solution
to the linearized equations governing a static equilibrium is spurious, that
each corresponds to a family of exact solutions, then the only solutions (ii)
are spherically symmetric, joining neighboring equilibria.

The decomposition into classes (i) and (ii) can be seen as follows. The set
of equations satisfied by $(\delta \rho, \delta \Phi, \delta v^a)$ are the 
perturbed mass conservation equation,
\be
\delta \left[ \partial_t \rho + \nabla_a(\rho v^a) \right] = 0,
\ee
the perturbed Euler equation,
\be
\delta \left[ (\partial_t + \pounds_v) v_a 
	+ \nabla_a(h - \half v^2 + \Phi) \right] = 0,
\ee
and the perturbed Poisson equation, $\delta\mbox{[Eq. (\ref{newton})]}$.

For a time-independent perturbation these equations take the form
\be 
\nabla_a(\rho \delta v^a) = 0, \label{continuity}
\ee
\be
\nabla_a(\delta h + \delta\Phi) = 0,  \label{pert_hydro}
\ee
and
\be
\nabla^2 \delta\Phi = 4\pi G\delta\rho,  \label{pert_newton}
\ee
where
\be
\delta h = \frac{\delta p}{\rho} = \frac{dp}{d\rho}\frac{\delta\rho}{\rho}. 
\label{pert_h}
\ee

Because Eq. (\ref{continuity}) for $\delta v^a$ decouples from 
Eqs. (\ref{pert_hydro}), (\ref{pert_newton}) and (\ref{pert_h}) for 
$(\delta \rho, \delta \Phi)$, any solution to Eqs. 
(\ref{continuity})-(\ref{pert_h}) is a superposition of a solution
$(0, 0, \delta v^a)$ and a solution $(\delta \rho, \delta \Phi, 0)$. This is
the claimed decomposition.

The theorem that any static self-gravitating perfect fluid is spherical 
implies that the solution $(\delta \rho, \delta \Phi, 0)$ is spherically
symmetric, to within the assumptions that the static perturbation
equations have no spurious solutions 
(``linearization stability'')\footnote{We are aware of a proof of this 
linearization stability for relativistic stars under assumptions on the 
equation of state that would not allow polytropes (\cite{ks80}).}.

Thus, under the assumption of linearization stability we have shown that 
all stationary non-radial perturbations of a spherical, isentropic star 
have $\delta \rho = \delta p = \delta \Phi = 0$ and a velocity field 
$\delta v^a$ that satisfies Eq. (\ref{continuity}).

A perturbation with axial parity has the form (\cite{jfs97}),
\be
\delta v^a = U(r) \epsilon^{abc}  \nabla_b 
			Y_l^m \nabla_c r, \label{form_ax}
\ee
and automatically satisfies Eq. (\ref{continuity}).

A perturbation with polar parity perturbation has the form,
\be
\delta v^a = \frac{W(r)}{r} Y_l^m \nabla^a r
			+ V(r) \nabla^a Y_l^m; \label{form_po}
\ee
and Eq. (\ref{continuity}) gives a relation between W and V,
\be
\frac{d}{d r} (r \rho W) - l(l+1) \rho V = 0.
\ee
These perturbations must satisfy the boundary conditions of regularity
at the center, $r=0$ and surface, $r=R$, of the star.  Also, the 
Lagrangian change in the pressure (defined in the next section) 
must vanish at the surface of the star.  These boundary conditions 
result in the requirement that
\be
W(0) = W(R) = 0;
\ee 
however, apart from this restriction, the radial functions $U(r)$ and $W(r)$ are
undetermined.

Thus, a spherical, isentropic, Newtonian star admits a class of zero
frequency convective fluid motions of the forms (\ref{form_ax}) and
(\ref{form_po}). Because they are stationary, these modes do not couple
to gravitational radiation.
\footnote{Note that for spherical stars, nonlinear
couplings invalidate the linear approximation after a time $t\sim R/\delta v$,
comparable to the time for a fluid element to move once around the star.
For nonzero angular velocity, the linear approximation is expected to be
valid for all times, if the amplitude is sufficiently small, roughly, if  
$|\delta v|< R\Omega$.}


\section{Rotating Isentropic Stars}

We consider perturbations of an isentropic Newtonian star, rotating
with uniform angular velocity $\Omega$.  No assumption of slow rotation
will be made until we turn to numerical computations in Sect. IV.  The
equilibrium of an axisymmetric, self-gravitating perfect fluid is
described by the gravitational potential $\Phi$, density $\rho$,
pressure $p$ and a 3-velocity
\be
v^a = \Omega \varphi^a,
\ee
where $\varphi^a$ is the rotational Killing vector field.

We will use a Lagrangian perturbation formalism (\cite{fs78a}) in which
perturbed quantities are described in terms of a Lagrangian
displacement vector $\xi^a$ that connects fluid elements in the
equilibrium and perturbed star.  The Eulerian change $\delta Q$ in a
quantity $Q$ is related to its Lagrangian change $\Delta Q$ by
\be
\Delta Q = \delta Q + \pounds_{\xi} Q,
\ee
with $\pounds_{\xi}$ the Lie derivative along $\xi^a$. 

The fluid perturbation is then determined by the displacement $\xi^a$:
\be
\Delta v^a = \partial_t \xi^a
\ee
\be
\frac{\Delta p}{\gamma p} = \frac{\Delta \rho}{\rho} = - \nabla_a \xi^a
\ee
Since the equilibrium spacetime is stationary and axisymmetric, we may
decompose our perturbations into modes of the form\footnote{We will always
choose $m \geq 0$ since the complex conjugate of an $m<0$ mode with frequency
$\sigma$ is an $m>0$ mode with frequency $-\sigma$.  Note that $\sigma$ is the 
frequency in an inertial frame.}
$e^{i(\sigma t + m \varphi)}$ .  The corresponding Eulerian changes are
\ba
\delta v^a  &=& i(\sigma + m \Omega) \xi^a \\
\delta \rho &=& - \nabla_a(\rho \xi^a) \\
\delta p    &=& \frac{dp}{d\rho} \delta \rho;
\ea
and the change in the gravitational potential is determined by
\be
\nabla^2 \delta \Phi = 4\pi G\delta \rho.
\ee


We can expand the perturbed fluid velocity, $\delta v^a$, in vector 
spherical harmonics (\cite{rw57}, see also \cite{th80}),
\be 
\delta v^a = \sum_{l=m}^{\infty} \left\{ \frac{1}{r}
	W_l Y_l^m \nabla^a r + V_l \nabla^a Y_l^m 
	- i U_l \epsilon^{abc} \nabla_b Y_l^m \nabla_c r 
	\right\} e^{i \sigma t},        \label{v_exp}
\ee
and examine the perturbed Euler equation.    

The Lagrangian perturbation of Euler's equation is
\ba
0 &=& \Delta [ (\partial_t + \pounds_v) v_a + 
	\nabla_a (h - \half v^2 + \Phi ) ]   \nonumber \\
  &=& (\partial_t + \pounds_v) \Delta  v_a 
	+ \nabla_a [ \Delta (h - \half v^2 + \Phi ) ], \label{pert_euler}
\ea
and its curl, which expresses the conservation of circulation for an 
isentropic star, is
\be
0 = q^a \equiv i(\sigma + m \Omega) \epsilon^{abc} \nabla_b \Delta v_c, \label{form_of_q}
\ee
or
\be
0 = q^a = i(\sigma + m \Omega) \epsilon^{abc} \nabla_b \delta v_c 
		+ \Omega \epsilon^{abc} \nabla_b ( \pounds_{\delta v} \varphi_c ).
\ee


Using the spherical harmonic expansion (\ref{v_exp}) of $\delta v^a$ we can 
write the components of $q^a$ as

\begin{eqnarray}
0=q^r &=& \frac{1}{r^2} \sum_{l=m}^{\infty} \left\{ \
	\vphantom{\frac{1}{2}}
	[(\sigma+m\Omega)l(l+1)- 2m\Omega ]U_lY_l^m 
	- 2 \Omega V_l[\sin\theta\partial_\theta Y_l^m 
				+ l(l+1)\cos\theta Y_l^m ]  \right. \nonumber  \\
    & & \left. \hspace{1in}
	\vphantom{\frac{1}{2}}
	\mbox{} + 2 \Omega W_l[\sin\theta\partial_\theta Y_l^m 
					+ 2 \cos\theta Y_l^m ] 
	 \ \right\} \ e^{i\sigma t},   \label{qr}
\end{eqnarray}

\begin{eqnarray}
0=q^{\theta} &=& \frac{1}{r^2 \sin\theta} \sum_{l=m}^{\infty} \left\{ \
	m (\sigma + m\Omega) \left(\partial_rV_l-\frac{W_l}{r}\right) Y_l^m 
	- 2\Omega\partial_rV_l\cos\theta\sin\theta\partial_\theta Y_l^m 
	\right. \nonumber\\
    	& & \left. \hspace{1in} 
	\mbox{} + 2\Omega m^2\frac{V_l}{r} Y_l^m 
	- 2\Omega\partial_rW_l\sin^2\theta Y_l^m 
	- 2m\Omega\partial_rU_l\cos\theta Y_l^m  \right. \nonumber\\
    	& & \left. \hspace{1in}
	\mbox{} + (\sigma + m\Omega)\partial_rU_l \sin\theta\partial_\theta Y_l^m
	+ 2m\Omega\frac{U_l}{r} \sin\theta\partial_\theta Y_l^m
	 \ \right\} \ e^{i\sigma t},    \label{qtheta}
\end{eqnarray}
and
\begin{eqnarray}
0=q^{\varphi} &=& \frac{i}{r^2 \sin^2\theta} \sum_{l=m}^{\infty} \left\{ \
	\vphantom{\frac{1}{2}}
	m (\sigma + m\Omega) \partial_rU_l Y_l^m
	- 2\Omega\partial_rU_l \cos\theta\sin\theta\partial_\theta Y_l^m 
	\right. \nonumber\\
	& & \left. \hspace{1in}
	\mbox{} + 2\Omega \frac{U_l}{r} [m^2-l(l+1)\sin^2\theta] Y_l^m   
	- 2m\Omega\partial_rV_l\cos\theta Y_l^m
	\right. \nonumber\\
	& & \left. \hspace{1in}
	\mbox{} + \left[ (\sigma + m\Omega)\left(
			\partial_rV_l-\frac{W_l}{r}\right) 
			+ 2m\Omega\frac{V_l}{r}
	  \right] \sin\theta\partial_\theta Y_l^m 
	 \ \right\} \ e^{i\sigma t}.    \label{qphi}
\end{eqnarray}
These components are not independent. The identity $\nabla_a q^a = 0$, which follows
from equation (\ref{form_of_q}), serves as a check on the right-hand sides of (\ref{qr}) 
- (\ref{qphi}).  


Let us rewrite these equations making use of the standard identities,
\begin{eqnarray}
\sin\theta\partial_\theta Y_l^m &=& l Q_{l+1} Y_{l+1}^m - (l+1) Q_l Y_{l-1}^m \\
\cos\theta Y_l^m &=& Q_{l+1} Y_{l+1}^m + Q_l Y_{l-1}^m 
\end{eqnarray}
where
\be
Q_l \equiv \left[ \frac{(l+m)(l-m)}{(2l-1)(2l+1)} \right]^{\half}. \label{Q_l}
\ee
Defining a dimensionless comoving frequency 
\be
\kappa \equiv \frac{(\sigma + m\Omega)}{\Omega},  \label{kappa}
\ee
we find that the $q^r=0$ equation becomes
\be
0  = \sum_{l=m}^{\infty} 
\begin{array}[t]{l}
\Biggl\{ \ [\half\kappa l(l+1) -m] U_lY_l^m \\
+ (W_l-lV_l)(l+2)Q_{l+1}Y_{l+1}^m - [W_l+(l+1)V_l](l-1)Q_lY_{l-1}^m
	 \ \Biggr\},                             \label{qr2}
\end{array}
\ee
$q^\theta =0$ becomes
\begin{eqnarray}
0 &=& \sum_{l=m}^{\infty} \left\{ \
- Q_{l+1}Q_{l+2} \left[ \vphantom{\frac{1}{2}}
lV'_l - W'_l \right] Y_{l+2}^m
- Q_{l+1} \left[ (m-\half\kappa l) U'_l - m l \frac{U_l}{r}\right] Y_{l+1}^m
	\right. \nonumber\\
	& & \left.
\mbox{} + \left[ 
\left( \half\kappa m +(l+1)Q_l^2 - l Q_{l+1}^2  \right) V'_l 
- \left(1- Q_l^2 - Q_{l+1}^2   \right) W'_l
- \half\kappa m \frac{W_l}{r} + m^2\frac{V_l}{r} 
\right] Y_l^m 
	\right. \nonumber\\
	& & \left.
\mbox{} - Q_l \left[ \left(m+\half\kappa (l+1)\vphantom{Q_l^2} \right) U'_l
+ m(l+1)\frac{U_l}{r} \right]  Y_{l-1}^m
	\right. \nonumber\\
	& & \left.
+ Q_{l-1} Q_l \left[ \vphantom{\frac{1}{2}}
(l+1)V'_l + W'_l \right] Y_{l-2}^m
	 \ \right\}                          \label{qtheta2}
\end{eqnarray}
and
$q^\varphi =0$ becomes
\begin{eqnarray}
0 &=& \sum_{l=m}^{\infty} \left\{ \
- l Q_{l+1}Q_{l+2} \left[ U'_l-(l+1) \frac{U_l}{r}\right] Y_{l+2}^m
	\right. \nonumber\\
	& & \left. 
\mbox{} + Q_{l+1} \left[ (\half\kappa l-m) V'_l + m l \frac{V_l}{r}
-\half\kappa l \frac{W_l}{r} \right]  Y_{l+1}^m
	\right. \nonumber\\
	& & \left. 
\mbox{} + \left[
  	\left( \half\kappa m +(l+1)Q_l^2 - l Q_{l+1}^2  \right) U'_l 
	+ \left( m^2 - l(l+1)\left(1- Q_l^2 - Q_{l+1}^2\right)\right)
	\frac{U_l}{r}
\right] Y_l^m
	\right. \nonumber\\
	& & \left. 
\mbox{} - Q_l \left[ \left(\half\kappa (l+1)+m \vphantom{Q_l^2}\right)V'_l
+ m(l+1)\frac{V_l}{r} - \half\kappa (l+1) \frac{W_l}{r} \right] Y_{l-1}^m
	\right. \nonumber\\
	& & \left. 
\mbox{} + (l+1)Q_{l-1}Q_l\left[ U'_l+l\frac{U_l}{r}\right] Y_{l-2}^m
	 \ \right\}                     \label{qphi2}
\end{eqnarray}
where $'\equiv\frac{d}{dr}$.

From this last form of the equations it is clear that the rotation of
the star mixes the axial and polar contributions to $\delta v^a$.  That
is, rotation mixes those terms in (\ref{v_exp}) whose limit as $\Omega
\rightarrow 0$ is axial with those terms in (\ref{v_exp}) whose limit
as $\Omega \rightarrow 0$ is polar.  It is also evident that the axial
contributions to $\delta v^a$ with $l$ even mix only with the odd $l$
polar contributions, and that the axial contributions with $l$ odd mix
only with the even $l$ polar contributions.  In addition, we prove in
appendix A that for non-axisymmetric modes the lowest value of $l$ that 
appears in the expansion of $\delta v^a$ is always $l=m$ (When $m=0$ this 
lowest value of $l$ is either 0 or 1.)  

Thus, we find two distinct classes of mixed, or hybrid, modes with definite 
behavior under parity.  This is to be expected because a rotating star is 
invariant under parity. Let us call a non-axisymmetric\footnote{When $m=0$
there exists a set of modes with parity $+1$ that may be designated as 
``axial-led hybrids'' since $\delta v^a$ receives contributions only from
axial terms with $l=1,3,5,\ldots$ and polar terms with $l=2,4,6,\ldots$.} 
mode an ``axial-led hybrid'' (or 
simply ``axial-hybrid'') if $\delta v^a$ receives contributions only from
\begin{center}
axial terms with $l \ = \ m, \ m+2, \ m+4,  \ \ldots$ and \\
polar terms with $l \ = \ m+1, \ m+3, \ m+5, \ \ldots$.
\end{center}
Such a mode has parity $(-1)^{m+1}$.

Similarly, we define a non-axisymmetric\footnote{When $m=0$ there exist two sets 
of modes that may be designated as ``polar-led hybrids.''  One set has parity
$-1$ and $\delta v^a$ receives contributions only from polar terms with 
$l=1,3,5,\ldots$ and axial terms with $l=2,4,6,\ldots$.  The other set (which
includes the radial oscillations) has parity $+1$ and $\delta v^a$ receives 
contributions only from polar terms with $l=0,2,4,\ldots$ and axial terms 
with $l=1,3,5,\ldots$.}
mode to be a ``polar-led hybrid'' (or ``polar-hybrid'') 
if $\delta v^a$ receives contributions only from
\begin{center}
polar terms with $l \ = \ m, \ m+2, \ m+4,  \ \ldots$ and \\
axial terms with $l \ = \ m+1, \ m+3, \ m+5,  \ \ldots$.
\end{center}
Such a mode has parity $(-1)^m$.


Let us rewrite the equations one last time using the orthogonality 
relation for spherical harmonics,
\be
\int Y_{l'}^{m'} Y_l^{\ast m} d\Omega = \delta_{ll'} \delta_{mm'},
\ee
where $d\Omega$ is the usual solid angle element.

From equation (\ref{qr2}) we find that $\int q^r Y_l^{\ast m} d\Omega = 0$ 
gives
\be
0 = 
[\half\kappa l(l+1) -m] U_l + (l+1)Q_l [W_{l-1}-(l-1)V_{l-1}] 
- lQ_{l+1} [W_{l+1}+(l+2)V_{l+1}]   \label{eq2}
\ee


Similarly, $\int q^{\theta} Y_l^{\ast m} d\Omega = 0$ gives
\begin{eqnarray}
0 &=& Q_lQ_{l-1} \{ (l-2)V'_{l-2}-W'_{l-2}\} + Q_l\left\{ [m-\half\kappa (l-1)]
  U'_{l-1}-m(l-1) \frac{U_{l-1}}{r}\right\}\nonumber\\
  &&\mbox{} + \left(1- Q_l^2 - Q_{l+1}^2   \right) W'_l
	- \left[\half\kappa m +(l+1)Q_l^2 - l Q_{l+1}^2  \right] V'_l 
	+ \half\kappa m \frac{W_l}{r} - m^2\frac{V_l}{r} \nonumber\\
  &&\mbox{} + Q_{l+1} \left\{ [m+\half\kappa (l+2)] U'_{l+1}+m(l+2)
  \frac{U_{l+1}}{r}\right\} \nonumber\\
  &&\mbox{} - Q_{l+2} Q_{l+1} \{ (l+3)V'_{l+2}+W'_{l+2}\} \label{eq3}
\end{eqnarray}
and $\int q^{\varphi} Y_l^{\ast m} d\Omega = 0$ gives
\begin{eqnarray}
0 &=& -(l-2)Q_lQ_{l-1} \left[ U'_{l-2}-(l-1) \frac{U_{l-2}}{r}\right] 
  + (l+3)Q_{l+2}Q_{l+1}\left[ U'_{l+2}+(l+2)\frac{U_{l+2}}{r}\right]\nonumber\\
  &&\mbox{} + \left\{ 
  	\left[\half\kappa m +(l+1)Q_l^2 - l Q_{l+1}^2  \right] U'_l 
	+ \left[ m^2 - l(l+1)\left(1- Q_l^2 - Q_{l+1}^2\right)\right] 
	\frac{U_l}{r}
	\right\} \nonumber\\
  &&\mbox{} + Q_l\left\{ [\half\kappa (l-1)-m] V'_{l-1} +m(l-1) \frac{V_{l-1}}{r}
  -\half\kappa (l-1) \frac{W_{l-1}}{r}\right\}\nonumber\\
  &&\mbox{} - Q_{l+1} \left\{ [\half\kappa (l+2)+m]V'_{l+1}+m(l+2)
  \frac{V_{l+1}}{r} - \half\kappa (l+2) \frac{W_{l+1}}{r}\right\}.  \label{eq4}
\end{eqnarray}


\section{Method of Solution}

In our numerical solution, we restrict consideration to slowly rotating 
stars, finding axial- and polar-led hybrids to lowest order in 
the angular velocity $\Omega$.  That is, we assume that perturbed quantities 
introduced above obey the following ordering in $\Omega$:
\be
\begin{array}{rrrr}
W_l \sim O(1),      & V_l   \sim O(1), & U_l \sim O(1), & \\
\delta \rho \sim O(\Omega), & \delta p \sim O(\Omega), & 
\delta \Phi \sim O(\Omega), & \sigma \sim O(\Omega).           \label{ordering}
\label{order}\end{array}
\ee
The $\Omega \rightarrow 0$ limit of such a perturbation is a sum of the
zero-frequency axial and polar perturbations considered in Sect. II. Note that, 
although the relative orders of $\delta \rho$ and $\delta v^a$
are physically meaningful, there is an arbitrariness in their absolute
order.  If $(\delta \rho, \delta v^a)$ is a solution to the linearized
equations, so is $(\Omega\delta \rho, \Omega\delta v^a)$.  We have chosen
the order (\ref{order}) to reflect the existence of well-defined,
nontrivial velocity perturbations of the spherical model.  Other authors (e.g., 
Lindblom and Ipser (1998)) adopt a convention in which $\delta v^a = O(\Omega)$ and
$\delta\rho = O(\Omega^2)$.

To lowest order, the equations governing these perturbations are the perturbed
Euler equations (\ref{eq2}) - (\ref{eq4}) and the perturbed mass conservation equation, 
(\ref{continuity}), which becomes
\be
rW'_l + \left( 1 + r \frac{\rho '}{\rho}\right) W_l - l(l+1) V_l = 0.  \label{eq1}
\ee

In addition, the perturbations must satisfy the boundary conditions of regularity
at the center of the star, $r=0$, regularity at the surface of 
the star, $r=R$, and the vanishing of the Lagrangian change in the pressure 
at the surface of the star,
\be
0 = \Delta p \equiv \delta p + \pounds_{\xi} p = \xi^r p' + \mbox{O($\Omega$)}.
\ee
Equations (\ref{eq2}) through (\ref{eq1}) are a system of ordinary differential 
equations for $W_{l'}(r)$, $V_{l'}(r)$ and $U_{l'}(r)$ (for all $l'$). Together with 
the boundary conditions, these equations form a non-linear eigenvalue problem for the 
parameter $\kappa$, where $\kappa\Omega$ is the mode frequency in the rotating frame.

To solve for the eigenvalues we proceed as follows.  We first ensure that the 
boundary conditions are automatically satisfied by expanding $W_{l'}(r)$, $V_{l'}(r)$ 
and $U_{l'}(r)$ (for all $l'$) in regular power series about the surface and center of the star.
Substituting these series into the differential equations results 
in a set of algebraic equations for the expansion coefficients. These algebraic equations 
may be solved for arbitrary values of $\kappa$ using standard matrix inversion methods.
For arbitrary values of $\kappa$, however, the series solutions about the center of 
the star will not necessarily agree with those about the surface of the star.  
The requirement that the series agree at some matching point, $0<r_0<R$, then 
becomes the condition that restricts the possible values of the eigenvalue, $\kappa_0$.  

The equilibrium solution $(\rho, \Phi)$ appears in the perturbation equations 
only through the quantity $(\rho'/\rho)$ in equation (\ref{eq1}). We begin by
writing the series expansion for this quantity about $r=0$ as
\be \left( \frac{\rho '}{\rho}\right) = \frac{1}{R}
\sum^\infty_{\stackrel{i=1}{\mbox{\tiny $i$ odd}}} 
\pi_i \left(\frac{r}{R}\right)^i,                     \label{rho_x}
\ee
and about $r=R$ as
\be \left(\frac{\rho '}{\rho}\right) = \frac{1}{R}
\sum^\infty_{k=-1} \tilde\pi_k \left(1-\frac{r}{R}\right)^k,  \label{rho_y}
\ee
where the $\pi_i$ and $\tilde\pi_k$ are determined from the equilibrium solution.

Because (\ref{eq2}) relates $U_l(r)$ algebraically to $W_{l\pm 1}(r)$ and  
$V_{l\pm 1}(r)$, we may eliminate $U_{l'}(r)$ (all $l'$) from 
(\ref{eq3}) and (\ref{eq4}).  We then need only work with one of equations 
(\ref{eq3}) or (\ref{eq4}) since the equations (\ref{eq2}) through 
(\ref{eq4}) are related by $\nabla_a q^a=0$.  

We next replace $\rho'/\rho$, $W_{l'}$, and $V_{l'}$ in equations (\ref{eq3})
or (\ref{eq4}) by their series expansions.  We eliminate the $U_{l'}(r)$
from either (\ref{eq3}) or (\ref{eq4}) and, again, substitute for the
$W_{l'}(r)$ and $V_{l'}(r)$.  Finally, we write down the matching condition
at the point $r_0$ equating the series expansions about $r=0$ to the
series expansions about $r=R$.  (Explicitly one equates (\ref{ax_x}) and (\ref{ax_y})
of Appendix B for axial-led modes or (\ref{po_x}) and (\ref{po_y}) for
polar-led modes).  The result is a linear algebraic system which
we may represent schematically as
\be Ax=0. \label{linalg} \ee
In this equation, $A$ is a matrix which depends non-linearly on the
parameter $\kappa$, and $x$ is a vector whose components are the
unknown coefficients in the series expansions for the $W_{l'}(r)$ and
$V_{l'}(r)$. In Appendix B, we explicitly present the equations making up
this algebraic system as well as the forms of the regular series
expansions for $W_{l'}(r)$ and $V_{l'}(r)$.

To satisfy equation (\ref{linalg}) we must find those values of
$\kappa$ for which the matrix $A$ is singular, i.e., we must find the
zeroes of the determinant of $A$.  We truncate the spherical harmonic
expansion of $\delta v^a$ at some maximum index $l_{\mbox{\tiny max}}$
and we truncate the radial series expansions about $r=0$ and $r=R$ at
some maximum powers $i_{\mbox{\tiny max}}$ and $k_{\mbox{\tiny max}}$, 
respectively.

The resulting finite matrix is band diagonal. To find the zeroes of its
determinant we use standard root finding techniques combined with
routines from the LAPACK linear algebra libraries (\cite{lapack}).  We find that
the eigenvalues, $\kappa_0$, computed in this manner converge quickly
as we increase $l_{\mbox{\tiny max}}$, $i_{\mbox{\tiny max}}$ and
$k_{\mbox{\tiny max}}$.

The eigenfunctions associated with these eigenvalues are determined by
the perturbation equations only up to normalization.  Given a
particular eigenvalue, we find its eigenfunction by replacing one of
the equations in the system (\ref{linalg}) with the normalization
condition that
\be
\begin{array}{ll}
V_m(r=R) = 1 & \mbox{for polar-hybrids, or that} \\
V_{m+1}(r=R) = 1 & \mbox{for axial-hybrids.} 
\end{array}
\ee
Since we have eliminated one of the rows of the singular matrix $A$ in favor
of this condition, the result is an algebraic system of the form
\be \tilde A x = b, \label{linalg2} \ee where $\tilde A$ is now a
non-singular matrix and $b$ is a known column vector.  We solve this
system for the vector $x$ using routines from LAPACK and reconstruct
the various series expansions from this solution vector of
coefficients.


\section{The Eigenvalues and Eigenfunctions}

We have computed the eigenvalues and eigenfunctions for uniform density stars
and for $n=1$ polytropes, models obeying the polytropic equation of state 
$p=K\rho^2$, where $K$ is a constant.  Our numerical solutions for the uniform 
density star agree with
the recent results of Lindblom and Ipser (1998) who find analytic solutions for 
the hybrid modes in rigidly rotating uniform density stars with arbitrary angular 
velocity - the Maclaurin spheroids.  Their calculation uses the two-potential 
formalism (\cite{im85}; and \cite{il90}) in which the equations for the perturbation modes 
are reformulated 
as coupled differential equations for a fluid potential, $\delta U$, and the gravitational 
potential, $\delta \Phi$. All of the perturbed fluid variables may be expressed in terms 
of these two potentials.  The analysis follows that of Bryan (1889) who found that the 
equations are separable in a non-standard spheroidal coordinate system.  

The Bryan/Lindblom-Ipser eigenfunctions $\delta U_0$ and $\delta \Phi_0$ turn out 
to be products of associated Legendre polynomials of their coordinates.  This 
simple form of their solutions
leads us to expect that our series solutions might also have a simple form - even
though their unusual spheroidal coordinates are rather complicated functions of
$r$ and $\theta$.  In fact, we do find that the modes of the uniform density
star have a particularly simple structure.  For any particular mode, both the angular 
and radial series expansions terminate at some finite indices $l_0$ and $i_0$ 
(or $k_0$).  That is, the spherical harmonic expansion (\ref{v_exp}) of $\delta v^a$ 
contains only terms with $m\leq l\leq l_0$ for this mode, and the coefficients of this 
expansion - the $W_l(r)$, $V_l(r)$ and $U_l(r)$ - are polynomials of order $i_0$.  
For all $l_0\geq m$ there exist a number of modes terminating at $l_0$.

In Tables \ref{ef_m1a} to \ref{ef_m2p} we present the functions 
$W_l(r)$, $V_l(r)$ and $U_l(r)$ for all of the
axial- and polar-led hybrids with $m=1$ and $m=2$ for a range of values of the terminating 
index $l_0$.  (See also Figure \ref{ef_mac}.)  For given values of $m>0$ and $l_0$ there 
are $l_0-m+1$ modes. (When $m=0$ there are $l_0$ modes. See equation (\ref{li_eq}) below.) 
We also find that the last term 
in the expansion (\ref{v_exp}), the term with $l=l_0$, is always axial for both types of 
hybrid modes.  This fact, together with the fact that the parity of the modes is,
\be
\pi = \Biggl\{ \begin{array}{lr}
(-1)^m     & \mbox{for polar-led hybrids\phm{,}} \\
(-1)^{m+1} & \mbox{for axial-led hybrids,} 
\end{array}
\ee
(for $m>0$) implies that $l_0-m+1$ must be even for polar-led modes and odd for axial-led
modes.

The fact that the various series terminate at $l_0$, $i_0$ and $k_0$ implies that 
Equations (\ref{linalg}) and (\ref{linalg2}) will be exact as long as we truncate 
the series at $l_{\mbox{\tiny max}}\geq l_0$, $i_{\mbox{\tiny max}}\geq i_0$ and 
$k_{\mbox{\tiny max}}\geq k_0$.

To find the eigenvalues of these modes we search the $\kappa$ axis for all of the 
zeroes of the determinant of the matrix $A$ in equation (\ref{linalg}). We begin by 
fixing $m$ and performing the search with $l_{\mbox{\tiny max}}=m$.  We then 
increase $l_{\mbox{\tiny max}}$ by 1 and repeat the search (and so on).  At any given 
value of $l_{\mbox{\tiny max}}$, the search finds all of the eigenvalues associated 
with the eigenfunctions terminating at $l_0\leq l_{\mbox{\tiny max}}$.

In Table \ref{ev_mac}, we present the eigenvalues $\kappa_0$ found by this method 
for the axial- and polar-led hybrid modes of uniform density stars for a range of values 
of $l_0$ and $m$. Observe that many of the eigenvalues, (marked with a $\ast$) satisfy 
the condition $\sigma(\sigma+m\Omega)<0$. [Recall that the mode frequency in an 
inertial frame is $\sigma=(\kappa_0-m)\Omega$]. The modes whose frequencies satisfy 
this condition are subject to a gravitational radiation driven instability in the 
absence of viscosity. The modes having $l_0=m>0$ (or $l_0=1$ for $m=0$) are the 
purely axial r-modes.  Their frequencies were found by Papalouizou and Pringle (1978) 
and are given by $\kappa_0 = 2/(m+1)$ (or $\kappa_0=0$ for $m=0$).  We find that 
there are no purely polar modes satisfying our assumptions (\ref{ordering}) in these 
stellar models.

We have compared these eigenvalues with those of Lindblom and Ipser (1998).  
To lowest non-trivial order in $\Omega$ their equation for 
the eigenvalue, $\kappa_0$, can be expressed in terms of associated
Legendre polynomials\footnote{The index $l$ used by Lindblom and Ipser is related to our
$l_0$ by $l=l_0+1$.  Our convention agrees with the usual labelling of the $l_0=m$ pure
axial modes.} (see Lindblom and Ipser's equation 6.4), as
\be
(4-\kappa_0^2) \frac{d}{d\kappa} P_{l_0+1}^m (\case{\kappa_0}{2}) 
- 2m P_{l_0+1}^m (\case{\kappa_0}{2}) = 0.                         \label{li_eq}
\ee
For given values of $m>0$ and $l_0$ this equation has $l_0-m+1$ roots (corresponding
to the number of distinct modes), which can easily be found numerically. (For $m=0$ 
there are $l_0$ roots.) For the range of values of $m$ and $l_0$ checked our eigenvalues 
agree with these to machine precision. (Compare our Table \ref{ev_mac} with Table 1 in 
\cite{li98}.)

We have also compared our eigenfunctions with those of Lindblom and Ipser. For a 
uniformly rotating, isentropic star, the fluid velocity perturbation, $\delta v^a$, 
is related (\cite{il90}) to $\delta U$ by 
\be
\nabla_a \delta U = - 
\left[ i\kappa\Omega g_{ab} + 2\nabla_b v_a \right] \delta v^b.
\ee
Since the $\varphi$ component of this equation is simply
\be
im \delta U = - \Omega r^2 \sin^2 \theta 
\left[  
\frac{2}{r}\delta v^r + 2\cot\theta\delta v^{\theta} + i\kappa\delta v^{\varphi}
\right],
\ee
it is straightforward numerically to construct this quantity from the components of 
our $\delta v^a$ and compare it with the analytic solutions for $\delta U$ given by 
Lindblom and Ipser (see their equation 7.2).  We have compared these solutions on a 
$20\times 40$ grid in the ($r-\theta$) plane and found that they agree (up to 
normalization) to better than $1$ part in $10^9$ for all cases checked.

Because of the use of the two-potential formalism and the unusual
coordinate system used in their analysis, the axial or polar hybrid
character of the Bryan/Lindblom-Ipser solutions is not obvious.  Nor is
it evident that these solutions have, as their $\Omega\rightarrow 0$
limit, the zero-frequency convective modes described in Sect. II.  The
comparison of their analytic results with our numerical work has served
the dual purpose of clarifying these properties of the solutions and of
testing the accuracy of our code.  The computational differences are
minor between the uniform density calculation and one in which the star
obeys a more realistic equation of state.  Thus, this testing gives us
confidence in the validity of our code for the polytrope calculation.
As a further check, we have written two independent codes and compared
the eigenvalues computed from each. One of these codes is based on the
set of equations described in Appendix B.  The other is based on the
set of second order equations that results from using the mass
conservation equation, (\ref{eq1}), to substitute for all the $V_l(r)$
in favor of the $W_l(r)$.

For the $n=1$ polytrope we will consider and, more generally, for any
isentropic equation of state, the purely axial r-modes are independent
of the equation of state.  In both isentropic and non-isentropic stars,
pure r-modes exist whose velocity field is, to lowest order in
$\Omega$, an axial vector field belonging to a single angular harmonic
(and restricted to harmonics with $l=m$ in the isentropic case).  The
frequency of such a mode is given (to order $\Omega$) by
$\kappa\Omega=(\sigma+m\Omega)=2m\Omega/l(l+1)$ (\cite{pp78}) and is
independent of the equation of state.  In isentropic stars, only those
modes having $l=m$ (or $l=1$ for $m=0$) exist, and for these modes the
eigenfunctions are also independent of the (isentropic) equation of
state\footnote{The only non-zero term in (\ref{v_exp}) for these modes
is the axial $l_0=m$ term with coefficient $U_m(r)=r^{m+1}$
(\cite{pea81}). The purely axial mode with $m=0$ has $l_0=1$ and radial
dependence $r^2$. This mode has zero frequency and corresponds to a
small uniform change in the angular velocity of the star.}.  This
independence of the equation of state occurs for the r-modes because
(to lowest order in $\Omega$) fluid elements move in surfaces of
constant $r$ (and thus in surfaces of constant density and pressure).
For the hybrid modes, however, fluid elements are not confined to
surfaces of constant $r$ and one would expect the eigenfrequencies and
eigenfunctions to depend on the equation of state.

Indeed, we find such a dependence.  The hybrid modes of the $n=1$ polytrope are not 
identical to those of the uniform density star. On the other hand, the modes do not appear 
to be very sensitive to the equation of state.  We have found that the character of the 
polytropic modes is similar to the modes of the uniform density star, except that the radial 
and angular series expansions do not terminate. For each eigenfunction in the uniform density 
star there is a corresponding eigenfunction in the polytrope with a slightly different 
eigenfrequency (See Table \ref{ev_poly}.)  For a given mode of the uniform density 
star, the series expansion (\ref{v_exp}) terminates at $l=l_0$.  For the corresponding 
polytrope mode, the expansion (\ref{v_exp}) does not terminate, but it does converge 
quickly.  The largest terms in (\ref{v_exp}) with
$l>l_0$ are more than an order of magnitude smaller than those with $l\leq l_0$ and they 
decrease rapidly as $l$ increases.  Thus, the terms that dominate the polytrope 
eigenfunctions are those that correspond to the non-zero terms in the corresponding 
uniform density eigenfunctions.  

In Figures \ref{ef_mac} and \ref{ef_poly} we display the coefficients $W_l(r)$, $V_l(r)$ and
$U_l(r)$ of the expansion (\ref{v_exp}) for the same $m=2$ axial-led hybrid mode in each 
stellar model.  For the uniform density star (Figure \ref{ef_mac}) the only non-zero 
coefficients for this mode are those with $l\leq l_0=4$.  These coefficients are presented
explicitly in Table \ref{ef_m2a} and are low order polynomials in $r$.  For the 
corresponding mode in the polytrope, we present in Figure \ref{ef_poly} the first seven
coefficients of the expansion (\ref{v_exp}).  Observe that those coefficients with $l\leq 4$ 
are similar to the corresponding functions in the uniform density mode and dominate the 
polytrope eigenfunction.  The coefficients with $4< l\leq 6$ are an order of magnitude 
smaller than the dominant coefficients and those with $l>6$ are smaller still. (Since they 
would be indistinguishable from the $(r/R)$ axis, we do not display the coefficients having
$l>6$ for this mode.)  

Just as the angular series expansion fails to terminate for the polytrope modes, so too do
the radial series expansions for the functions $W_l(r)$, $V_l(r)$ and $U_l(r)$.  We have seen
that in the uniform density star these functions are polynomials in $r$ (Tables \ref{ef_m1a}
through \ref{ef_m2p}).  In the polytropic star, the radial series do not terminate and we are
required to work with both sets of radial series expansions - those about the center of the star
and those about its surface - in order to represent the functions accurately everywhere inside 
the star.

In Figures \ref{fig3} through \ref{fig11} we compare corresponding functions from each type
of star.  For example, Figures \ref{fig3}, \ref{fig4}, and \ref{fig5} show the functions 
$W_l(r)$, $V_l(r)$ and $U_l(r)$ (respectively) for $l\leq 6$ for a particular $m=1$ polar-led hybrid mode.  
In the uniform density star this mode has eigenvalue $\kappa_0=1.509941$, and in the polytrope 
it has eigenvalue $\kappa_0=1.412999$.  The only non-zero functions in the uniform density mode 
are those with 
$l\leq l_0=2$ and they are simple polynomials in $r$ (see Table \ref{ef_m1p}).  Observe that 
these functions are similar, but not identical to, their counterparts in the polytrope mode,
which have been constructed from their radial series expansions about $r=0$ and $r=R$ (with
matching point $r_0=0.5R$).  Again, note the convergence with increasing $l$ of the 
polytrope eigenfunction.  The mode is dominated by the terms with $l\leq 2$ and those with 
$l>2$ decrease rapidly with $l$. (The $l=5$ and $l=6$ coefficients are virtually 
indistinguishable from the $(r/R)$ axis.)

Because the polytrope eigenfunctions are dominated by their $l\leq l_0$ terms, the
eigenvalue search with $l_{\mbox{\tiny max}}=l_0$ will find the associated eigenvalues
approximately.  We compute these approximate eigenvalues of the polytrope modes using 
the same search technique as for the uniform density star.  We then increase 
$l_{\mbox{\tiny max}}$ and search near one of the approximate eigenvalues for a corrected
value, iterating this procedure until the eigenvalue converges to the desired accuracy.
We present the eigenvalues found by this method in Table \ref{ev_poly}.

As a further comparison between the mode eigenvalues in the polytropic star and those in 
the uniform density star we have modelled a sequence of ``intermediate'' stars. By
multiplying the expansions (\ref{rho_x}) and (\ref{rho_y}) for $(\rho'/\rho)$ by a 
scaling factor, $\epsilon\in [0,1]$, we can simulate a continuous sequence of stellar
models connecting the uniform density star ($\epsilon=0$) to the polytrope
($\epsilon=1$).  We find that an eigenvalue in the uniform density star varies smoothly 
as function of $\epsilon$ to the corresponding eigenvalue in the polytrope.


\section{The Effects of Dissipation}

The effects of gravitational radiation and viscosity on the pure $l_0=m$ r-modes
have already been studied by a number of authors. (Lindblom et al 1998, \cite{o98}, 
Andersson et al 1998, \cite{ks98}, \cite{lmo99}) All of these modes are unstable to 
gravational radiation reaction, and for some of them this instability strongly dominates 
viscous damping. We now consider the effects of dissipation on the axial- and 
polar-hybrid modes.

To estimate the timescales associated with viscous damping and gravitational radiation
reaction we follow the methods used for the $l_0=m$ modes (Lindblom et al 1998, see also 
\cite{il91}).  When the energy radiated per cycle is small compared to the energy of
the mode, the imaginary part of the mode frequency is accurately approximated by the
expression
\be
\frac{1}{\tau} = - \frac{1}{2E} \frac{dE}{dt},   \label{tau}
\ee
where $E$ is the energy of the mode as measured in the rotating frame,
\be
E = \frac{1}{2} \int \left[ 
\rho \delta v^a \delta v^{\ast}_a + \left( \frac{\delta p}{\rho}+\delta\Phi\right) 
\delta\rho^{\ast}
\right] d^3 x .                                                       \label{E}
\ee
The rate of change of this energy due to dissipation by viscosity and gravitational 
radiation is,
\begin{eqnarray}
\frac{dE}{dt} &=& - \int \left( 
2\eta\delta\sigma^{ab}\delta\sigma_{ab}^{\ast} +\zeta \delta\theta\delta\theta^{\ast}
\right) \nonumber \\
 & & -\sigma (\sigma + m\Omega) \sum_{l\geq 2} N_l \sigma^{2l} \left(
\left|\delta D_{lm}\right|^2   + \left|\delta J_{lm}\right|^2
\right).                                                                \label{dEdt}
\end{eqnarray}
The first term in (\ref{dEdt}) represents dissipation due to shear viscosity, where 
the shear, $\delta\sigma_{ab}$, of the perturbation is
\be
\delta\sigma_{ab} = \half\left( 
\nabla_a \delta v_b + \nabla_b \delta v_a - \case{2}{3} g_{ab} \nabla_c \delta v^c
\right),
\ee
and the coefficient of shear viscosity for hot neutron-star matter is 
(\cite{cl87}; \cite{s89})
\be
\eta = 2\times 10^{18} 
\left(\frac{\rho}{10^{15}\mbox{g}\!\cdot\!\mbox{cm}^{-3}}\right)^{\case{9}{4}}
\left(\frac{10^9K}{T}\right)^2 \ 
\mbox{g}\!\cdot\!\mbox{cm}^{-1}\!\cdot\!\mbox{s}^{-1}.                 \label{eta}
\ee

The second term in (\ref{dEdt}) represents dissipation due to bulk viscosity, where 
the expansion, $\delta\theta$, of the perturbation is
\be
\delta\theta = \nabla_c \delta v^c
\ee
and the bulk viscosity coefficient for hot neutron star matter is 
(\cite{cl87}; \cite{s89})
\be
\zeta = 6\times 10^{25} 
\left(\frac{1\mbox{Hz}}{\sigma + m\Omega}\right)^2
\left(\frac{\rho}{10^{15}\mbox{g}\!\cdot\!\mbox{cm}^{-3}}\right)^2
\left(\frac{T}{10^9K}\right)^6 \ 
\mbox{g}\!\cdot\!\mbox{cm}^{-1}\!\cdot\!\mbox{s}^{-1}.                 \label{zeta}
\ee

The third term in (\ref{dEdt}) represents dissipation due to gravitational 
radiation, with coupling constant
\be
N_l = \frac{4\pi G}{c^{2l+1}}\frac{(l+1)(l+2)}{l(l-1)[(2l+1)!!]^2}.
\ee
The mass, $\delta D_{lm}$, and current, $\delta J_{lm}$, multipole moments of the 
perturbation are given by (\cite{th80}, Lindblom et al 1998)
\be
\delta D_{lm} = \int \delta\rho r^l Y_l^{\ast m} d^3 x,                \label{D}
\ee 
and
\be
\delta J_{lm} = \frac{2}{c} \left(\frac{l}{l+1}\right)^{\half} \int r^l 
\left( \rho\delta v_a + \delta\rho v_a \right) Y^{a,B\ast}_{lm} d^3 x    \label{J}
\ee
where $Y^{a,B}_{lm}$ is the magnetic type vector spherical harmonic (\cite{th80}) given by,
\be
Y^{a,B}_{lm} = \frac{r}{\sqrt{l(l+1)}} \epsilon^{abc} \nabla_b Y_l^{m} \nabla_c r.
\ee

To lowest order in $\Omega$, the energy (\ref{E}) of the hybrid modes is positive definite.  Their
stability is therefore determined by the sign of the right hand side of equation (\ref{dEdt}).
We have seen that many of the hybrid modes have frequencies satisfying 
$\sigma(\sigma+m\Omega)<0$.  This makes the third term in (\ref{dEdt}) positive, implying
that gravitational radiation reaction tends to drive these modes unstable. 
(\cite{ch70}; \cite{fs78b}; \cite{jf78}) To determine
the actual stability of these modes, we must evaluate the various dissipative terms in
(\ref{dEdt}).

We first substitute for $\delta v^a$ the spherical harmonic expansion (\ref{v_exp}) and 
use the orthogonality relations for vector spherical harmonics (\cite{th80}) to perform the
angular integrals.  The energy of the modes in the rotating frame then becomes
\be
E = \sum^{\infty}_{l=m} \half \int_0^R \rho 
\left[ W_l^2 + l(l+1)V_l^2 + l(l+1)U_l^2 \right] dr.
\ee

To calculate the dissipation due to gravitational radiation reaction we must evaluate
the multipole moments (\ref{D}) and (\ref{J}).  To lowest order in $\Omega$ the mass
multipole moments vanish and the current multipole moments are given by 
\be
\delta J_{lm} = \frac{2l}{c} \int_0^R \rho r^{l+1} U_l dr.  \label{J_int}
\ee

To calculate the dissipation due to bulk viscosity we must evaluate the expansion,
$\delta\theta = \nabla_c \delta v^c$, of the perturbation.  For
uniform density stars this quantity vanishes identically by the mass conservation
equation (\ref{continuity}).  For the $l_0=m$, pure axial modes
the expansion, again, vanishes identically, regardless of the equation of state.  
To compute the bulk viscosity of these modes it is necessary to work to 
higher order in $\Omega$ (\cite{aks98}, \cite{lmo99}).  On the other hand, for the new 
hybrid modes in which we are interested, the expansion of the fluid perturbation 
is non-zero in the slowly rotating polytropic stars. After substituting for 
$\delta v^a$ its series expansion and performing the angular 
integrals, the bulk viscosity contribution to (\ref{dEdt}) becomes
\be
\left( \frac{dE}{dt} \right)_B = - \sum^{\infty}_{l=m} \int_0^R 
\frac{\zeta}{r^2} \left[ r W'_l + W_l - l(l+1)V_l \right]^2 dr
\ee
In a similar manner, the contribution to (\ref{dEdt}) from shear viscosity becomes
\be
\left( \frac{dE}{dt} \right)_S = - \sum^{\infty}_{l=m} \int_0^R  \frac{2\eta}{r^2}
\begin{array}[t]{l}
\Biggl\{
\case{2}{3} \left[ r^3 \left(\frac{W_l}{r^2}\right)'\right]^2 
+ \half l(l+1) W_l^2 
+ \half l(l+1)\left[ r^3 \left(\frac{V_l}{r^2}\right)'\right]^2  \\
+ \case{1}{3} l(l+1)(2l^2+2l-3) V_l^2 
+  l(l+1) W_l \left[ r^5 \left(\frac{V_l}{r^4}\right)'\right] 
+ \case{2}{3} l(l+1) V_l \left(rW_l\right)' \\
+ \half l(l+1) \left[ r^3 \left(\frac{U_l}{r^2}\right)'\right]^2 
+ \half l(l+1)(l^2+l-2) U_l^2 
\Biggr\} dr.
\end{array}
\ee

Given a numerical solution for one of the hybrid mode eigenfunctions, these radial 
integrals can be performed numerically. The resulting contributions to (\ref{dEdt})
also depend on the angular velocity and temperature of the star.  Let us express
the imaginary part of the hybrid mode frequency (\ref{tau}) as,
\be
\frac{1}{\tau} = \frac{1}{\tilde \tau_S} \left( \frac{10^9 K}{T} \right)^2
+ \frac{1}{\tilde \tau_B} \left( \frac{T}{10^9 K} \right)^6 
\left( \frac{\pi G \bar{\rho}}{\Omega^2} \right) 
+ \sum_{l\geq 2} \frac{1}{\tilde \tau_l}
\left( \frac{\Omega^2}{\pi G \bar{\rho}} \right)^{l+1}, \label{tau2}
\ee
where $\bar{\rho}$ is average density of the star. (Compare this expression to the 
corresponding expression in Lindblom et al. (1998) - their equation (22) - for the 
$l_0=m$ pure axial modes.)  

The bulk viscosity term in this equation is stronger by a factor $\Omega^{-4}$ than 
that for the $l_0=m$ pure axial modes. This is because the expansion $\delta\theta$ 
of the hybrid mode is nonzero to lowest order in 
$\Omega$ for the polytropic star, whereas it is order $\Omega^2$ for the
pure axial modes.  This implies that the damping due to bulk viscosity 
will be much stronger for the hybrid modes than for the pure axial modes
in slowly rotating stars.

Note that the contribution to (\ref{tau2}) from gravitational
radiation reaction consists of a sum over all the values of $l$ with
a non-vanishing current multipole.  This sum is, of course, dominated by 
the lowest contributing multipole.

In Tables \ref{times_m1a} to \ref{times_m2p} we present the timescales for 
these various dissipative effects in the uniform density and polytropic stellar 
models that we have been considering with $R=12.57\mbox{km}$ and $M=1.4M_{\sun}$.  
For the reasons discussed above, we do not present bulk viscosity timescales for the 
uniform density star.

Given the form of their eigenfunctions, it seems reasonable to expect that some of 
the unstable hybrid modes might grow on a timescale which is comparable to that of the 
pure $l_0=m$ r-modes. For example, the $m=2$ axial-led hybrids all have 
$U_2(r)\neq 0$ (see, for example, Figures \ref{ef_mac} and \ref{ef_poly}). By equation 
(\ref{J_int}), this leads one to 
expect a non-zero current quadrupole moment $\delta J_{22}$, and this is the multipole 
moment that dominates the gravitational radiation in the r-modes. Upon closer 
inspection, however, one finds that this is not the case.  In fact, we find that all of 
the multipoles $\delta J_{lm}$ vanish (or nearly vanish) for $l<l_0$, where $l_0$ is 
the largest value of $l$ contributing a dominant term to the expansion (\ref{v_exp}) 
of $\delta v^a$.  

In the uniform density star, these multipoles vanish identically.  Consider, for example, 
the $m=2$, $l_0=4$ axial-hybrid with eigenvalue $\kappa=0.466901$. (See Table \ref{ef_m2a}
and Figure \ref{fig6}) 
For this mode, $U_2\propto (7x^3-9x^5)$, where $x=(r/R)$. By equation (\ref{J_int}), 
we then find that
\be
\delta J_{22} \propto \int_0^1 x^3 (7x^3-9x^5) dx \equiv 0,
\ee
and that $\delta J_{42}$ is the only non-zero radiation multipole.  In general, the
only non-zero multipole for an axial- or polar-hybrid mode in the uniform density star
is $\delta J_{l_0\, m}$. 

That this should be the case is not obvious from the form of our eigenfunctions.  However, 
Lindblom and Ipser's (1998) analytic solutions provide an explanation.  Their equations (7.1) 
and (7.3) reveal that the perturbed gravitational potential, $\delta \Phi$, is a pure
spherical harmonic to lowest order in $\Omega$.  In particular,
\be
\delta \Phi \propto Y_{l_0+1}^m.
\ee
This implies that the only non-zero current multipole is $\delta J_{l_0\, m}$.

We find a similar result for the polytropic star. Because of the similarity between the 
modes in the polytrope and the modes in the uniform density star, we find that although
the lower $l$ current multipoles do not vanish identically, they very nearly vanish and
the radiation is dominated by higher $l$ multipoles.

The fastest growth times we find in the hybrid modes are of order $10^4$ seconds (at
$10^9K$ and $\Omega=\sqrt{\pi G\bar{\rho}}$).  Thus, the spin-down of a newly
formed neutron star will be dominated by the $l_0=m=2$ mode with contributions from
the $l_0=m$ pure axial modes with $2\leq m\lesssim 10$ and from the fastest growing 
hybrid modes.


\section{Discussion}

There is substantial uncertainty in the cooling rate of neutron
stars, with rapid cooling expected if stars have a quark interior or
core, or a kaon or pion condensate. Madsen (1998) suggests that an
observation of a young neutron star with a rotation period below 
$5-10\mbox{ms}$ would be evidence for a quark interior; but even without 
rapid cooling, the uncertainty in the superfluid transition temperature 
would allow a superfluid to form at about $10^{10} K$, killing the instability.  
The nonaxisymmetric instability has been expected not to play a role in old
neutron stars spun up by accretion, because of the high shear viscosity
associated with an expected temperature $\leq 10^7 K$; but even this is
not certain (\cite{akst98}).
 
    An extension of our numerical method to find modes of rapidly rotating 
Newtonian models and slowly rotating relativistic models appears
feasible. Work is in progress to understand the way in which the modes
join the r- and g- modes of stars that are not isentropic (\cite{aea99}).


\acknowledgements

We wish to thank Sharon Morsink and Lee Lindblom for numerous 
discussions and for helpful comments on an earlier draft of this 
paper.  We also thank Nils Andersson, Kostas Kokkotas, Yasufumi 
Kojima, Bernard Schutz and Nick Stergioulas for helpful discussions 
and for sharing related work in progress with us. We are grateful to 
the AEI, Potsdam, for generous hospitality during the first part of 
this research. This work was supported in part by NSF Grant PHY-9507740.


\appendix


\section{The character of the modes of rotating isentropic stars}

For an equilibrium model that is axisymmetric and invariant under parity,
one can resolve any degeneracy in the perturbation spectrum to make each discrete
mode an eigenstate of parity with angular dependence $e^{im\varphi}$. The following 
theorem holds.

\begin{thm}{Let $(\delta\rho, \delta v^a)$ with $\delta v^a\neq 0$ be a discrete normal
mode of a uniformly rotating stellar model obeying a one-parameter equation of state.
Then the decomposition of the mode into spherical harmonics $Y_l^m$ (i.e., into $(l, m)$ 
representations of the rotation group about its center of mass) has $l=m$ as the lowest 
contributing value of $l$, when $m\neq 0$; and has 0 or 1 as the lowest contributing 
value of $l$, when $m=0$.}
\end{thm}

In Sect. III, we designate non-axisymmetric modes with parity $(-1)^m$ ``polar-led hybrids'', 
and non-axisymmetric modes with parity $(-1)^{m+1}$ ``axial-led hybrids,'' and briefly
discuss the $m=0$ case.

Note that the theorem holds for p-modes as well as for the rotational modes that are 
our main concern.  A p-mode is determined by its density perturbation and is therefore 
dominantly polar in character regardless of its parity.  For a rotational mode, however, 
the lowest $l$ term in its velocity perturbation is at least comparable in magnitude 
to the other contributing terms.  

We prove the theorem separately for each parity class.

\subsection{Axial-Led Hybrids with $m>0$}

Let $l$ be the smallest value of $l'$ for which $U_{l'}\neq 0$ in the spherical
harmonic expansion (\ref{v_exp}) of the perturbed velocity field $\delta v^a$.  
The axial parity of $\delta v^a$, $(-1)^{l+1}$, and the vanishing of $Y_l^m$ for $l<m$ 
implies $l\geq m$.  That the mode is axial-led means $W_{l'}=0$ and 
 $V_{l'}=0$ for $l'\leq l$.   We show by contradiction that $l=m$.

Suppose $l\geq m+1$. From equation (\ref{eq2}), $\int q^r Y_l^{\ast m} d\Omega = 0$, 
we have
\be
[\half\kappa l(l+1) -m] U_l = lQ_{l+1} [W_{l+1}+(l+2)V_{l+1}],
\ee
and from equation (\ref{eq3}) with $l$ replaced by $l-1$, 
$\int q^{\theta} Y_{l-1}^{\ast m}d\Omega = 0$, we have
\be
Q_{l+1} \left[ (l+2)V'_{l+1} + W'_{l+1} \right] = 
\left\{ [m+\half\kappa (l+1)] U'_l + m(l+1) \frac{U_l}{r} \right\}.
\ee
These two equations, together imply that 
\[
U'_l + \frac{l}{r} U_l = 0,
\]
or 
\[
U_l = K r^{-l},
\]
which is singular at $r=0$.

\subsection{Axial-Led Hybrids with $m=0$}

Let $m=0$ and let $l$ be the smallest value of $l'$ for which $U_{l'}\neq 0$ in the spherical
harmonic expansion (\ref{v_exp}) of the perturbed velocity field $\delta v^a$.  
Since $\nabla_a Y_0^0=0$, the mode vanishes unless $l\geq 1$. That the mode is 
axial-led means $W_{l'}=0$ and $V_{l'}=0$ for $l'\leq l$.  We show by 
contradiction that $l=1$.

Suppose $l \geq 2$. Then $\int q^{\varphi} Y_{l-2}^{\ast 0} d\Omega = 0$ becomes,
\be
U'_l + \frac{l}{r} U_l = 0,
\ee
or 
\[
U_l = K r^{-l},
\]
which is singular at $r=0$.

\subsection{Polar-Led Hybrids with $m>0$}

Let $l$ be the smallest value of $l'$ for which $W_{l'}\neq 0$ or $V_{l'}\neq 0$ in the spherical
harmonic expansion (\ref{v_exp}) of the perturbed velocity field $\delta v^a$.  
The polar parity of $\delta v^a$, $(-1)^l$, and the vanishing of $Y_l^m$ for $l<m$ 
implies $l\geq m$.  That the mode is polar-led means $U_{l'}=0$ for $l'\leq l$.   
We show by contradiction that $l=m$.

Suppose $l \geq m+1$. Then $\int q^r Y_{l-1}^{\ast m} d\Omega = 0$ becomes
\be
W_l+(l+1)V_l = 0,
\ee
and $\int q^{\varphi} Y_{l-1}^{\ast m} d\Omega = 0$ becomes,
\be
0 = 
\begin{array}[t]{l}
- \left\{[\half\kappa (l+1)+m] V'_l + m(l+1)\frac{V_l}{r} 
- \half\kappa (l+1)\frac{W_l}{r} 
\right\} \\
+ (l+2) Q_{l+1} \left[ U'_{l+1} + (l+1) \frac{U_{l+1}}{r} \right].
\end{array}
\ee
These two equations, together imply that
\[
- [\half\kappa (l+1)+m] \left[V'_l  + (l+1) \frac{V_l}{r} \right]
+ (l+2) Q_{l+1} \left[ U'_{l+1} + (l+1) \frac{U_{l+1}}{r} \right]
= 0,
\]
or
\[
- [\half\kappa (l+1)+m] V_l + (l+2) Q_{l+1} U_{l+1} = K r^{-{(l+1)}},
\]
which is singular at $r=0$.

\subsection{Polar-Led Hybrids with $m=0$}

Let $m=0$ and let $l$ be the smallest value of $l'$ for which $W_{l'}\neq 0$ and 
$V_{l'}\neq 0$ in the spherical harmonic expansion (\ref{v_exp}) of the perturbed 
velocity field $\delta v^a$. When $l=0$ the mode is automatically polar-led; thus we need
only consider the case $l\geq 1$. That the mode is polar-led means $U_{l'}=0$ for 
$l'\leq l$.  We show by contradiction that $l=1$.

Suppose $l \geq 2$. Then $\int q^r Y_{l-1}^{\ast 0} d\Omega = 0$ becomes
\be
W_l+(l+1)V_l = 0,
\ee
and $\int q^{\varphi} Y_{l-1}^{\ast 0} d\Omega = 0$ becomes,
\be
- \half\kappa (l+1)\left[V'_l - \frac{W_l}{r} \right]
+ (l+2) Q_{l+1} \left[ U'_{l+1} + (l+1) \frac{U_{l+1}}{r} \right].
= 0
\ee
These two equations, together imply that
\[
- \half\kappa (l+1) \left[V'_l  + (l+1) \frac{V_l}{r} \right]
+ (l+2) Q_{l+1} \left[ U'_{l+1} + (l+1) \frac{U_{l+1}}{r} \right] = 0,
\]
or
\[
- \half\kappa (l+1) V_l + (l+2) Q_{l+1} U_{l+1} = K r^{-{(l+1)}},
\]
which is singular at $r=0$.


\section{The algebraic equations governing the hybrid modes to lowest order in $\Omega$.}

\noindent In this appendix, we make use of the following definitions:
\begin{eqnarray}
a_l &\equiv& \half\kappa m +(l+1)Q_l^2 - l Q_{l+1}^2 \\
b_l &\equiv& m^2 - l(l+1)\left(1- Q_l^2 - Q_{l+1}^2\right) \\
c_l &\equiv& \half\kappa l(l+1) - m 
\end{eqnarray}
For reference, we repeat the definitions (\ref{Q_l}) and (\ref{kappa}):
\begin{eqnarray}
\kappa  &\equiv& \frac{(\sigma + m\Omega)}{\Omega} \\
Q_l     &\equiv& \left[ \frac{(l+m)(l-m)}{(2l-1)(2l+1)} \right]^{\half}
\end{eqnarray}

\subsection{Axial Hybrids}

\noindent For $l=m,\:m+2,\:m+4,\ldots$ the regular series expansions\footnote{We present 
the form of the series expansions for $U_l(r)$ for reference; however, we do not need 
these series since we eliminate the $U_l(r)$ using equation (\ref{eq2}).} about
the center of the star, $r=0$, are
\begin{mathletters}
\begin{eqnarray}
W_{m+j+1}(r) &=& \left(\frac{r}{R}\right)^{m+j} 
\sum^\infty_{\stackrel{i=1}{\mbox{\tiny $i$ odd}}} 
w_{j+1,i} \left(\frac{r}{R}\right)^i                    
\\
V_{m+j+1}(r) &=& \left(\frac{r}{R}\right)^{m+j} 
\sum^\infty_{\stackrel{i=1}{\mbox{\tiny $i$ odd}}} 
v_{j+1,i} \left(\frac{r}{R}\right)^i                     
\\
U_{m+j}(r) &=& \left(\frac{r}{R}\right)^{m+j} 
\sum^\infty_{\stackrel{i=1}{\mbox{\tiny $i$ odd}}} 
u_{j,i} \left(\frac{r}{R}\right)^i
\end{eqnarray} \label{ax_x}
\end{mathletters}
where $j=0,2,4,\ldots$.

\noindent The regular series expansions about $r=R$, which satisfy the boundary 
condition $\Delta p = 0$ are
\begin{mathletters}
\begin{eqnarray}
W_{m+j+1}(r) &=& \sum^\infty_{k=1} \tilde w_{j+1,k} \left(1-\frac{r}{R}\right)^k
\\
V_{m+j+1}(r) &=& \sum^\infty_{k=0} \tilde v_{j+1,k} \left(1-\frac{r}{R}\right)^k
\\
U_{m+j}(r) &=& \sum^\infty_{k=0} \tilde u_{j,k} \left(1-\frac{r}{R}\right)^k
\end{eqnarray} \label{ax_y}
\end{mathletters}
where $j=0,2,4,\ldots$.

\noindent
These series expansions must agree in the interior of the star. We impose the matching 
condition that the series (\ref{ax_x}) truncated at $i_{\mbox{\tiny max}}$ be equal at the 
point $r=r_0$ to the corresponding series (\ref{ax_y}) truncated at $k_{\mbox{\tiny max}}$. 
That is,
\begin{mathletters}
\begin{eqnarray}
0 &=& \left(\frac{r_0}{R}\right)^{m+j} 
\sum^{i_{\mbox{\tiny max}}}_{\stackrel{i=1}{\mbox{\tiny $i$ odd}}} 
w_{j+1,i} \left(\frac{r_0}{R}\right)^i                    
- \sum^{k_{\mbox{\tiny max}}}_{k=1} \tilde w_{j+1,k} \left(1-\frac{r_0}{R}\right)^k
\label{ax_match_a}
\\
0 &=& \left(\frac{r_0}{R}\right)^{m+j} 
\sum^{i_{\mbox{\tiny max}}}_{\stackrel{i=1}{\mbox{\tiny $i$ odd}}} 
v_{j+1,i} \left(\frac{r_0}{R}\right)^i                     
- \sum^{k_{\mbox{\tiny max}}}_{k=0} \tilde v_{j+1,k} \left(1-\frac{r_0}{R}\right)^k
\label{ax_match_b}
\end{eqnarray} \label{ax_match}
\end{mathletters}

\noindent 
When we substitute (\ref{ax_x}) and (\ref{rho_x}) into (\ref{eq1}), the 
coefficient of $(r/R)^{m+j+i}$ in the resulting equation is
\be
0 = (m+j+i+1)w_{j+1,i} + \sum^{i-2}_{\stackrel{s=1}{\mbox{\tiny $s$ odd}}} 
\pi_s \, w_{j+1,i-s-1} - (m+j+1)(m+j+2)v_{j+1,i}        \label{ax_1}
\ee

\noindent Similarly, when we substitute (\ref{ax_y}) and (\ref{rho_y}) into 
(\ref{eq1}), the coefficient of $[1-(r/R)]^k$ in the resulting equation is
\be
0 = (k+1) \left[\tilde w_{j+1,k} - \tilde w_{j+1,k+1} \right]
+ \sum^k_{s=0} \left(\tilde\pi_{s-1} - \tilde\pi_{s-2} \right) \tilde w_{j+1,k-s+1} 
- (m+j+2)(m+j+1)\tilde v_{j+1,k}      \label{ax_2}
\ee
where we have defined $\tilde \pi_{-2} \equiv 0 \equiv \tilde w_{j+1,0}$.

\noindent When we use (\ref{eq2}) to eliminate the $U_l(r)$ from (\ref{eq4}) 
and then substitute for the $W_{l\pm 1}(r)$ and $V_{l\pm 1}(r)$ using 
(\ref{ax_x}), the coefficient of $(r/R)^{m+j+i}$ in the resulting equation is
\begin{eqnarray}
0 &=& (i+1)(m+j-2)(m+j-1)Q_{m+j}Q_{m+j-1}Q_{m+j-2}c_{m+j}c_{m+j+2}
\nonumber\\
&& \hspace{2.0in}
\mbox{} \times \Biggl[ w_{j-3,i+4} - (m+j-3) v_{j-3,i+4} \Biggr]
\nonumber\\
&&\mbox{} - Q_{m+j} c_{m+j+2} \Biggl\{
(i+1)(m+j-2)^2Q^2_{m+j-1}c_{m+j}+\half\kappa (m+j-1)c_{m+j-2}c_{m+j}
\nonumber\\
&& \hspace{1.0in}
\mbox{} + (m+j+1) \left[(m+j+i)a_{m+j}+b_{m+j}\right] c_{m+j-2}
\Biggr\} w_{j-1,i+2}
\nonumber\\
&&\mbox{} + Q_{m+j}c_{m+j+2}  \Biggl\{ 
\left[\half\kappa (m+j-1)(m+j+i)-(i+1)m\right] c_{m+j-2}c_{m+j}
\nonumber\\
&& \hspace{1.0in}
\mbox{} + (m+j+1)(m+j-1) \left[(m+j+i)a_{m+j}+b_{m+j} \right] c_{m+j-2}
\nonumber\\
&& \hspace{1.0in}
\mbox{} -  (i+1)(m+j)(m+j-2)^2Q^2_{m+j-1}c_{m+j}
\Biggr\} v_{j-1,i+2}
\nonumber\\
&&\mbox{} + Q_{m+j+1}c_{m+j-2} \Biggl\{ 
\half\kappa (m+j+2) c_{m+j}c_{m+j+2} 
+ (m+j)\left[(m+j+i)a_{m+j}+b_{m+j}\right] c_{m+j+2} 
\nonumber\\
&& \hspace{1.0in}
\mbox{} - (2m+2j+i+2)(m+j+3)^2Q^2_{m+j+2}c_{m+j}
\Biggr\} w_{j+1,i} 
\nonumber\\
&&\mbox{} + Q_{m+j+1}c_{m+j-2} \Biggl\{
(m+j+2)(m+j)\left[(m+j+i)a_{m+j}+b_{m+j}\right]c_{m+j+2}
\nonumber\\
&& \hspace{1.0in}
\mbox{} - \left[\half\kappa (m+j+2)(m+j+i)+m(2m+2j+i+2)\right] c_{m+j}c_{m+j+2}
\nonumber\\
&& \hspace{1.0in}
\mbox{} + (2m+2j+i+2)(m+j+3)^2(m+j+1)Q^2_{m+j+2}c_{m+j}
\Biggr\} v_{j+1,i} 
\nonumber\\
&&\mbox{} + (2m+2j+i+2)(m+j+3)(m+j+2)Q_{m+j+3}Q_{m+j+2}Q_{m+j+1} c_{m+j-2}c_{m+j}
\nonumber\\
&& \hspace{2.0in}
\mbox{} \times \Biggl[ w_{j+3,i-2} + (m+j+4) v_{j+3,i-2} \Biggr]        \label{ax_3} 
\end{eqnarray}

\noindent When we use (\ref{eq2}) to eliminate the $U_l(r)$ from (\ref{eq4}) and 
then substitute for the $W_{l\pm 1}(r)$ and $V_{l\pm 1}(r)$ using (\ref{ax_y}), 
the coefficient of $[1-(r/R)]^k$ in the resulting equation is
\begin{eqnarray}
0 &=& - (m+j-k-1)(m+j-1)(m+j-2)Q_{m+j}Q_{m+j-1}Q_{m+j-2}c_{m+j}c_{m+j+2}
\nonumber\\
&& \hspace{2.0in}
\mbox{} \times \Biggl[ \tilde w_{j-3,k} - (m+j-3) \tilde v_{j-3,k} \Biggr]
\nonumber\\
&&\mbox{} - (k+1)(m+j-1)(m+j-2)Q_{m+j}Q_{m+j-1}Q_{m+j-2}c_{m+j}c_{m+j+2}
\nonumber\\
&& \hspace{2.0in}
\mbox{} \times \Biggl[ \tilde w_{j-3,k+1} - (m+j-3) \tilde v_{j-3,k+1} \Biggr]
\nonumber\\
&&\mbox{} + Q_{m+j}c_{m+j+2} \Biggl\{
(m+j-k-1)(m+j-2)^2Q^2_{m+j-1}c_{m+j} - \half\kappa(m+j-1)c_{m+j-2}c_{m+j}
\nonumber\\
&& \hspace{1.0in}
\mbox{} -(m+j+1)\left(b_{m+j}+ka_{m+j}\right)c_{m+j-2}
\Biggr\} \tilde w_{j-1,k} 
\nonumber\\
&&\mbox{} + (k+1)Q_{m+j}c_{m+j+2} \Biggl\{
(m+j-2)^2Q^2_{m+j-1}c_{m+j} + (m+j+1)a_{m+j}c_{m+j-2}
\Biggr\} \tilde w_{j-1,k+1} 
\nonumber\\
&&\mbox{} + Q_{m+j}c_{m+j+2} \Biggl\{
(m+j-k-1)(m+j-2)^2(m+j)Q^2_{m+j-1}c_{m+j}
\nonumber\\
&& \hspace{1.0in}
\mbox{} + \left[\half\kappa k(m+j-1)+m(m+j-k-1)\right] c_{m+j-2}c_{m+j}
\nonumber\\
&& \hspace{1.0in}
\mbox{} + (m+j+1)(m+j-1)\left(b_{m+j}+ka_{m+j}\right) c_{m+j-2}
\Biggr\} \tilde v_{j-1,k} 
\nonumber\\
&&\mbox{} + (k+1)Q_{m+j}c_{m+j+2} \Biggl\{
(m+j)(m+j-2)^2Q^2_{m+j-1}c_{m+j} 
\nonumber\\
&& \hspace{1.0in}
\mbox{} + \left[m-\half\kappa (m+j-1)\right] c_{m+j-2}c_{m+j}
\nonumber\\
&& \hspace{1.0in}
\mbox{} -(m+j+1)(m+j-1)a_{m+j}c_{m+j-2}
\Biggr\} \tilde v_{j-1,k+1}
\nonumber\\
&&\mbox{} + Q_{m+j+1}c_{m+j-2} \Biggl\{
(m+j)\left(b_{m+j}+ka_{m+j}\right)c_{m+j+2} + \half\kappa (m+j+2)c_{m+j}c_{m+j+2} 
\nonumber\\
&& \hspace{1.0in}
\mbox{} - (m+j+k+2)(m+j+3)^2Q^2_{m+j+2}c_{m+j}
\Biggr\} \tilde w_{j+1,k} 
\nonumber\\
&&\mbox{} + (k+1)Q_{m+j+1}c_{m+j-2} \Biggl\{
-(m+j)a_{m+j}c_{m+j+2}+(m+j+3)^2Q^2_{m+j+2}c_{m+j}
\Biggr\} \tilde w_{j+1,k+1} 
\nonumber\\
&&\mbox{} + Q_{m+j+1}c_{m+j-2} \Biggl\{
(m+j+2)(m+j)\left(b_{m+j}+ka_{m+j}\right)c_{m+j+2}
\nonumber\\
&& \hspace{1.0in}
\mbox{} - \left[m(m+j+k+2)+\half\kappa k(m+j+2)\right] c_{m+j}c_{m+j+2}
\nonumber\\
&& \hspace{1.0in}
\mbox{} +(m+j+k+2)(m+j+3)^2(m+j+1)Q^2_{m+j+2}c_{m+j}
\Biggr\} \tilde v_{j+1,k} 
\nonumber\\
&&\mbox{} + (k+1)Q_{m+j+1}c_{m+j-2} \Biggl\{
-(m+j+2)(m+j)a_{m+j}c_{m+j+2}
\nonumber\\
&& \hspace{1.0in}
\mbox{}+\left[\half\kappa (m+j+2)+m\right] c_{m+j}c_{m+j+2}
\nonumber\\
&& \hspace{1.0in}
\mbox{} -(m+j+3)^2(m+j+1)Q^2_{m+j+2}c_{m+j}
\Biggr\} \tilde v_{j+1,k+1}
\nonumber\\
&&\mbox{} + (m+j+k+2)(m+j+3)(m+j+2)Q_{m+j+3}Q_{m+j+2}Q_{m+j+1}c_{m+j-2}c_{m+j}
\nonumber\\
&& \hspace{2.0in}
\mbox{} \times \Biggl[ \tilde w_{j+3,k} + (m+j+4) \tilde v_{j+3,k} \Biggr]
\nonumber\\
&&\mbox{} - (k+1)(m+j+3)(m+j+2)Q_{m+j+3}Q_{m+j+2}Q_{m+j+1}c_{m+j-2}c_{m+j}
\nonumber\\
&& \hspace{2.0in}
\mbox{} \times \Biggl[ \tilde w_{j+3,k+1} + (m+j+4) \tilde v_{j+3,k+1} \Biggr]
\label{ax_4}
\end{eqnarray}

The equations (\ref{ax_match}) through (\ref{ax_4}) make up the algebraic system 
(\ref{linalg}) for eigenvalues of the axial-led hybrid modes.  One truncates the angular 
and radial series expansions at indices $j_{\mbox{\tiny max}}$, $i_{\mbox{\tiny max}}$ and 
$k_{\mbox{\tiny max}}$ and constructs the matrix $A$ by keeping the appropriate number of 
equations for the number of unknown coefficients $w_{j+1,i}$, $v_{j+1,i}$, 
$\tilde w_{j+1,k}$ and $\tilde v_{j+1,k}$. In following this procedure, however, one must 
be aware of the following subtlety in the equations.

For each $q\equiv j+i$ the set of equations
\[
\begin{array}{ll}
\mbox{(\ref{ax_1})} & \mbox{with} \ \ i=1 \ \ \mbox{and} \ \ j=q-1, \ \ \mbox{and} \\
\mbox{(\ref{ax_3})} & \mbox{for all} \ \ i=1,3,\ldots,q \ \ \mbox{with} \ \ j=q-i
\end{array}
\]
can be shown to be linearly dependent for arbitrary $\kappa$ and for any equilibrium
stellar model. For example, taking the simplest case of $q=1$, one can show that equation 
(\ref{ax_1}) with $i=1$ and $j=0$ becomes
\[
0 = (m+2) \left[w_{1,1} - (m+1) v_{1,1} \right] 
\]
while equation (\ref{ax_3}) with $i=1$ and $j=0$ becomes
\[
0 = 
\begin{array}[t]{l}
Q_{m+1}c_{m-2} \Biggl\{
\half\kappa(m+2)c_m c_{m+2} 
+ m\left[(m+1)a_m + b_m \right]
- (2m+3)(m+3)^2 Q^2_{m+2}c_m
\Biggr\} \\
\times \left[w_{1,1} - (m+1) v_{1,1} \right].
\end{array}
\]

This problem can be solved by eliminating one of these equations from the subset for 
each $q$ (for example, equation (\ref{ax_3}) with $i=1$).  Thus, to properly construct the 
algebraic system (\ref{linalg}) we use, for all $j=0,2,\ldots,j_{\mbox{\tiny max}}$, 
the equations
\[
\begin{array}{ll}
\mbox{(\ref{ax_match_a})} & \\
\mbox{(\ref{ax_match_b})} & \\
\mbox{(\ref{ax_1})} & \mbox{with} \ \ i=1,3,\ldots,i_{\mbox{\tiny max}} \\ 
\mbox{(\ref{ax_2})} & \mbox{with} \ \ k=0,1,\ldots,k_{\mbox{\tiny max}}-1 \\ 
\mbox{(\ref{ax_3})} & \mbox{with} \ \ i=3,5,\ldots,i_{\mbox{\tiny max}} \\
\mbox{(\ref{ax_4})} & \mbox{with} \ \ k=0,1,\ldots,k_{\mbox{\tiny max}}-1. \\ 
\end{array}
\]

\subsection{Polar Hybrids}

\noindent For $l=m,\:m+2,\:m+4,\ldots$ the regular series expansions\footnote{We present 
the form of the series expansions for $U_l(r)$ for reference; however, we do not need 
these series since we eliminate the $U_l(r)$ using equation (\ref{eq2}).} about
the center of the star, $r=0$, are
\begin{mathletters}
\begin{eqnarray}
W_{m+j}(r) &=& \left(\frac{r}{R}\right)^{m+j} 
\sum^\infty_{\stackrel{i=0}{\mbox{\tiny $i$ even}}} 
w_{j,i} \left(\frac{r}{R}\right)^i
\\
V_{m+j}(r) &=& \left(\frac{r}{R}\right)^{m+j} 
\sum^\infty_{\stackrel{i=0}{\mbox{\tiny $i$ even}}} 
v_{j,i} \left(\frac{r}{R}\right)^i                    
\\
U_{m+j+1}(r) &=& \left(\frac{r}{R}\right)^{m+j} 
\sum^\infty_{\stackrel{i=2}{\mbox{\tiny $i$ even}}} 
u_{j+1,i} \left(\frac{r}{R}\right)^i
\end{eqnarray} \label{po_x}
\end{mathletters}
where $j=0,2,4,\ldots$.

\noindent The regular series expansions about $r=R$, which satisfy the boundary 
condition $\Delta p = 0$ are
\begin{mathletters}
\begin{eqnarray}
W_{m+j}(r) &=& \sum^\infty_{k=1} \tilde w_{j,k} \left(1-\frac{r}{R}\right)^k
\\
V_{m+j}(r) &=& \sum^\infty_{k=0} \tilde v_{j,k} \left(1-\frac{r}{R}\right)^k
\\
U_{m+j+1}(r) &=& \sum^\infty_{k=0} \tilde u_{j+1,k} \left(1-\frac{r}{R}\right)^k
\end{eqnarray}  \label{po_y}
\end{mathletters}
where $j=0,2,4,\ldots$.

\noindent
These series expansions must agree in the interior of the star. We impose the matching 
condition that the series (\ref{po_x}) truncated at $i_{\mbox{\tiny max}}$ be equal at the 
point $r=r_0$ to the corresponding series (\ref{po_y}) truncated at $k_{\mbox{\tiny max}}$. 
That is,
\begin{mathletters}
\begin{eqnarray}
0 &=& \left(\frac{r_0}{R}\right)^{m+j} 
\sum^{i_{\mbox{\tiny max}}}_{\stackrel{i=0}{\mbox{\tiny $i$ even}}} 
w_{j,i} \left(\frac{r_0}{R}\right)^i                    
- \sum^{k_{\mbox{\tiny max}}}_{k=1} \tilde w_{j,k} \left(1-\frac{r_0}{R}\right)^k
\label{po_match_a}
\\
0 &=& \left(\frac{r_0}{R}\right)^{m+j} 
\sum^{i_{\mbox{\tiny max}}}_{\stackrel{i=0}{\mbox{\tiny $i$ even}}} 
v_{j,i} \left(\frac{r_0}{R}\right)^i                     
- \sum^{k_{\mbox{\tiny max}}}_{k=0} \tilde v_{j,k} \left(1-\frac{r_0}{R}\right)^k
\label{po_match_b}
\end{eqnarray} \label{po_match}
\end{mathletters}

\noindent 
When we substitute (\ref{po_x}) and (\ref{rho_x}) into (\ref{eq1}), the 
coefficient of $(r/R)^{m+j+i}$ in the resulting equation is
\be
0 = (m+j+i+1)w_{j,i} + \sum^{i-2}_{\stackrel{s=0}{\mbox{\tiny $s$ even}}} 
\pi_{s+1} \, w_{j,i-s-2} - (m+j)(m+j+1)v_{j,i}        \label{po_1}
\ee

\noindent Similarly, when we substitute (\ref{po_y}) and (\ref{rho_y}) into 
(\ref{eq1}), the coefficient of $[1-(r/R)]^k$ in the resulting equation is
\be
0 = (k+1) \left[\tilde w_{j,k} - \tilde w_{j,k+1} \right]
+ \sum^k_{s=0} \left(\tilde\pi_{s-1} - \tilde\pi_{s-2} \right) \tilde w_{j,k-s+1} 
- (m+j)(m+j+1)\tilde v_{j,k}                     \label{po_2}
\ee
where we have defined $\tilde \pi_{-2} \equiv 0 \equiv \tilde w_{j,0}$.

\noindent When we use (\ref{eq2}) to eliminate the $U_l(r)$ from (\ref{eq3}) 
and then substitute for the $W_{l\pm 1}(r)$ and $V_{l\pm 1}(r)$ using 
(\ref{po_x}), the coefficient of $(r/R)^{m+j+i}$ in the resulting equation is
\begin{eqnarray}
0 &=& - im(m+j-1)Q_{m+j}Q_{m+j-1} c_{m+j+1} 
\Biggl[ w_{j-2, i+2} - (m+j-2)v_{j-2, i+2} \Biggr]
\nonumber\\
&&\mbox{} + \Biggl\{ 
(m+j-1)Q^2_{m+j}[(i+1)m-\half\kappa(m+j-1)(m+j+i)] c_{m+j+1}
\nonumber\\
& & \hspace{0.5in}
\mbox{} + \left[
(m+j+i)\left(1- Q_{m+j}^2 - Q_{m+j+1}^2\right)+\half\kappa m
\right] c_{m+j-1} c_{m+j+1} 
\nonumber\\
& & \hspace{0.5in}
\mbox{} - 
(m+j+2)Q^2_{m+j+1} [m(2m+2j+i+2)+\half\kappa (m+j+2)(m+j+i)] c_{m+j-1}
\Biggr\}  w_{j,i} 
\nonumber\\
&&\mbox{} + \Biggl\{ 
(m+j-1)(m+j+1)Q^2_{m+j}[(i+1)m-\half\kappa (m+j-1)(m+j+i)] c_{m+j+1}
\nonumber\\
&& \hspace{0.5in}
\mbox{} - \left[ m^2 + (m+j+i)a_{m+j} \right]  c_{m+j-1} c_{m+j+1} 
\nonumber\\
&& \hspace{0.5in}
\mbox{} + \left.
(m+j)(m+j+2)Q^2_{m+j+1} \vphantom{\frac{1}{2}} \right.
\nonumber\\
&& \hspace{1.5in}
\mbox{} \times \left[m(2m+2j+i+2)+ \half\kappa (m+j+2)(m+j+i)\right]  c_{m+j-1}
\Biggr\}  v_{j,i}
\nonumber\\
&&\mbox{} + Q_{m+j+2}Q_{m+j+1}\left[m(m+j+i)+m(m+j+1)(2m+2j+i+2)\right] c_{m+j-1}
\nonumber\\
&& \hspace{1.5in}
\mbox{} \times \Biggl[ w_{j+2, i-2} + (m+j+3)v_{j+2, i-2} \Biggr]
\label{po_3}
\end{eqnarray}

\noindent When we use (\ref{eq2}) to eliminate the $U_l(r)$ from (\ref{eq3}) and 
then substitute for the $W_{l\pm 1}(r)$ and $V_{l\pm 1}(r)$ using (\ref{po_y}), 
the coefficient of $[1-(r/R)]^k$ in the resulting equation is
\begin{eqnarray}
0 &=& m(m+j-1)(m+j-k) Q_{m+j}Q_{m+j-1} c_{m+j+1} 
\Biggl[ \tilde w_{j-2,k} - (m+j-2) \tilde v_{j-2, k} \Biggr]
\nonumber\\
&&\mbox{} + (k+1)m(m+j-1) Q_{m+j}Q_{m+j-1} c_{m+j+1}
\Biggl[ \tilde w_{j-2,k+1} - (m+j-2) \tilde v_{j-2, k+1} \Biggr]
\nonumber\\
&&\mbox{} + \Biggl\{ 
- (m+j-1) Q^2_{m+j} [(\half\kappa k + m)(m+j-1)-km] c_{m+j+1}
\nonumber\\
&& \hspace{0.5in}
\mbox{} + \left[
\half\kappa m + k\left(1- Q_{m+j}^2 - Q_{m+j+1}^2\right) \right] c_{m+j-1} c_{m+j+1}
\nonumber\\
&& \hspace{0.5in}
\mbox{} - (m+j+2) Q^2_{m+j+1}[(\half\kappa k+m)(m+j+2)+km] c_{m+j-1}
\Biggr\} \tilde w_{j,k} 
\nonumber\\
&&\mbox{} - (k+1) \Biggl\{ 
(m+j-1)Q^2_{m+j}[m-\half\kappa (m+j-1)] c_{m+j+1}
\nonumber\\
&& \hspace{1.0in}
\mbox{} + \left(1- Q_{m+j}^2 - Q_{m+j+1}^2\right) c_{m+j-1} c_{m+j+1}
\nonumber\\
&& \hspace{1.0in}
\mbox{} - (m+j+2) Q^2_{m+j+1}[m+\half\kappa (m+j+2)] c_{m+j-1}
\Biggr\} \tilde w_{j,k+1} 
\nonumber\\
&&\mbox{} + \Biggl\{ 
-(m+j-1)(m+j+1) Q^2_{m+j}[(\half\kappa k +m)(m+j-1)-km] c_{m+j+1}
\nonumber\\
&& \hspace{0.5in}
\mbox{} - \left( m^2+ka_{m+j}\right) c_{m+j-1} c_{m+j+1}
\nonumber\\
&& \hspace{0.5in}
\mbox{} + (m+j)(m+j+2) Q^2_{m+j+1}[(\half\kappa k +m)(m+j+2)+km] c_{m+j-1}
\Biggr\} \tilde v_{j,k} 
\nonumber\\
&&\mbox{} + (k+1)\Biggl\{ 
- (m+j-1)(m+j+1) Q^2_{m+j}[m-\half\kappa (m+j-1)] c_{m+j+1}
\nonumber\\
&& \hspace{1.0in}
\mbox{} + a_{m+j} c_{m+j-1} c_{m+j+1}
\nonumber\\
&& \hspace{1.0in}
\mbox{} - (m+j)(m+j+2) Q^2_{m+j+1}[m+\half\kappa (m+j+2)] c_{m+j-1}
\Biggr\} \tilde v_{j,k+1}
\nonumber\\
&&\mbox{} + m(m+j+2)(m+j+k+1) Q_{m+j+2}Q_{m+j+1} c_{m+j-1}
\Biggl[ \tilde w_{j+2,k} + (m+j+3) \tilde v_{j+2,k} \Biggr]
\nonumber\\
&&\mbox{} - (k+1)m(m+j+2) Q_{m+j+2}Q_{m+j+1} c_{m+j-1}
\Biggl[ \tilde w_{j+2,k+1} + (m+j+3) \tilde v_{j+2,k+1} \Biggr]
\label{po_4}
\end{eqnarray}

The equations (\ref{po_match}) through (\ref{po_4}) make up the algebraic system 
(\ref{linalg}) for eigenvalues of the polar-led hybrid modes.  As in the case of the axial-led
hybrids, one truncates the angular and radial series expansions at indices 
$j_{\mbox{\tiny max}}$, $i_{\mbox{\tiny max}}$ and $k_{\mbox{\tiny max}}$ and constructs the 
matrix $A$ by keeping the appropriate number of equations for the number of unknown 
coefficients $w_{j,i}$, $v_{j,i}$, $\tilde w_{j,k}$ and $\tilde v_{j,k}$. 

We, again, find that certain subsets of these equations are linearly dependent for
arbitrary $\kappa$ and for any equilibrium stellar model.  For all $j$, it can be shown that 
both equation (\ref{po_1}) with $i=0$ and equation (\ref{po_3}) with $i=0$ are proportional to 
\[
0 = \left[w_{j,0} - (m+j) v_{j,0} \right].
\]

This problem can, again, be solved by eliminating, for example, equation (\ref{po_3}) 
with $i=0$ for all $j$.  Thus, to properly construct the algebraic system (\ref{linalg}) 
we use, for all $j=0,2,\ldots,j_{\mbox{\tiny max}}$, 
the equations
\[
\begin{array}{ll}
\mbox{(\ref{po_match_a})} & \\
\mbox{(\ref{po_match_b})} & \\
\mbox{(\ref{po_1})} & \mbox{with} \ \ i=0,2,\ldots,i_{\mbox{\tiny max}} \\ 
\mbox{(\ref{po_2})} & \mbox{with} \ \ k=0,1,\ldots,k_{\mbox{\tiny max}}-1 \\ 
\mbox{(\ref{po_3})} & \mbox{with} \ \ i=2,4,\ldots,i_{\mbox{\tiny max}} \\
\mbox{(\ref{po_4})} & \mbox{with} \ \ k=0,1,\ldots,k_{\mbox{\tiny max}}-1. \\ 
\end{array}
\]



\newpage

\figcaption{All of the non-zero coefficients $W_l(r)$, $V_l(r)$, $U_l(r)$ of the 
spherical harmonic expansion (\ref{v_exp}) for a particular $m=2$ axial-led hybrid mode of
the uniform density star.  The mode has eigenvalue $\kappa_0 = -0.763337$.  Note 
that the largest value of $l$ that appears in the expansion (\ref{v_exp}) is 
$l_0=4$ and that the functions $W_l(r)$, $V_l(r)$ and $U_l(r)$ are low order 
polynomials in $(r/R)$. (See Table \ref{ef_m2a}.) The mode is normalized so that 
$V_2(r=R)=1$.
 \label{ef_mac}}

\figcaption{The coefficients $W_l(r)$, $V_l(r)$, $U_l(r)$ with $l\leq 6$ of the 
spherical harmonic expansion (\ref{v_exp}) for a particular $m=2$ axial-led hybrid mode of the 
polytropic star.  This is the polytrope mode that corresponds to the uniform density 
mode displayed in Figure \ref{ef_mac}. For the polytrope the mode has eigenvalue 
$\kappa_0 = -1.025883$.  The expansion (\ref{v_exp}) converges rapidly with 
increasing $l$ and is dominated by the terms with $2\leq l\leq 4$, i.e., by the terms 
corresponding to those which are non-zero for the uniform density mode.  Observe that 
the coefficients shown with $4<l\leq 6$ are an order of magnitude smaller than those 
with $2\leq l\leq 4$. Those with $l>6$ are smaller still and are not displayed here.  
The mode is, again, normalized so that $V_2(r=R)=1$. 
 \label{ef_poly}}

\figcaption{The functions $W_l(r)$ with $l\leq 6$ for a particular $m=1$ polar-led 
hybrid mode. For the uniform density star this mode has eigenvalue
$\kappa_0=1.509941$ and $W_1=-x+x^3$ ($x=r/R$) is the only non-vanishing 
$W_l(r)$ (see Table \ref{ef_m1p}). The corresponding mode of the polytropic star 
has eigenvalue $\kappa_0=1.412999$.  Observe that $W_1(r)$ for the polytrope, which 
has been constructed from its power series expansions about $r=0$ and $r=R$, is similar, 
though not identical, to the corresponding $W_1(r)$ for the uniform density star.  
Observe also that the functions $W_l(r)$ with $l>1$ for the polytrope are more than an order of 
magnitude smaller than $W_1(r)$ and become smaller with increasing $l$. ($W_5(r)$ is 
virtually indistinguishable from the $(r/R)$ axis.)
\label{fig3}}

\figcaption{The functions $V_l(r)$ with $l\leq 6$ for the same mode as in Figure \ref{fig3}.
 \label{fig4}}

\figcaption{The functions $U_l(r)$ with $l\leq 6$ for the same mode as in Figure \ref{fig3}.
 \label{fig5}}

\figcaption{The functions $U_l(r)$ with $l\leq 7$ for a particular $m=2$ axial-led hybrid mode.  
For the uniform density star this mode has eigenvalue $\kappa_0=0.466901$ and $U_2(r)$ and 
$U_4(r)$ are the only non-vanishing $U_l(r)$. (See Table \ref{ef_m2a} for their explicit 
forms.) The corresponding mode of the polytropic star has eigenvalue $\kappa_0=0.517337$. 
Observe that $U_2(r)$ and $U_4(r)$ for the polytrope, which have been constructed from their 
power series expansions about $r=0$ and $r=R$, are similar, though not identical, to the 
corresponding functions for the uniform density star.  Observe also that the $U_6(r)$ 
is more than an order of magnitude smaller than $U_2(r)$ and $U_4(r)$.  
 \label{fig6}}

\figcaption{The functions $W_l(r)$ with $l\leq 7$ for the same mode as in Figure \ref{fig6}.
 \label{fig7}}

\figcaption{The functions $V_l(r)$ with $l\leq 7$ for the same mode as in Figure \ref{fig6}.
 \label{fig8}}

\figcaption{The functions $U_l(r)$ with $l\leq 8$ for a particular $m=2$ axial-led hybrid mode.  
For the uniform density star this mode has eigenvalue $\kappa_0=0.359536$ and $U_2(r)$, 
$U_4(r)$ and $U_6(r)$ are the only non-vanishing $U_l(r)$. (See Table \ref{ef_m2a} for their 
explicit forms.) The corresponding mode of the polytropic star has eigenvalue 
$\kappa_0=0.421678$.  Observe that $U_2(r)$, $U_4(r)$ and $U_6(r)$ for the polytrope, 
which have been constructed from their power series expansions about $r=0$ and $r=R$, are 
similar, though not identical, to the corresponding functions for the uniform density star.  
Observe also that $U_8(r)$ is more than an order of magnitude smaller than $U_2(r)$, $U_4(r)$ 
and $U_6(r)$.  
 \label{fig9}}

\figcaption{The functions $W_l(r)$ with $l\leq 8$ for the same mode as in Figure \ref{fig9}.
 \label{fig10}}

\figcaption{The functions $V_l(r)$ with $l\leq 8$ for the same mode as in Figure \ref{fig9}.
 \label{fig11}}



\resizebox{\hsize}{9.5in}{\includegraphics{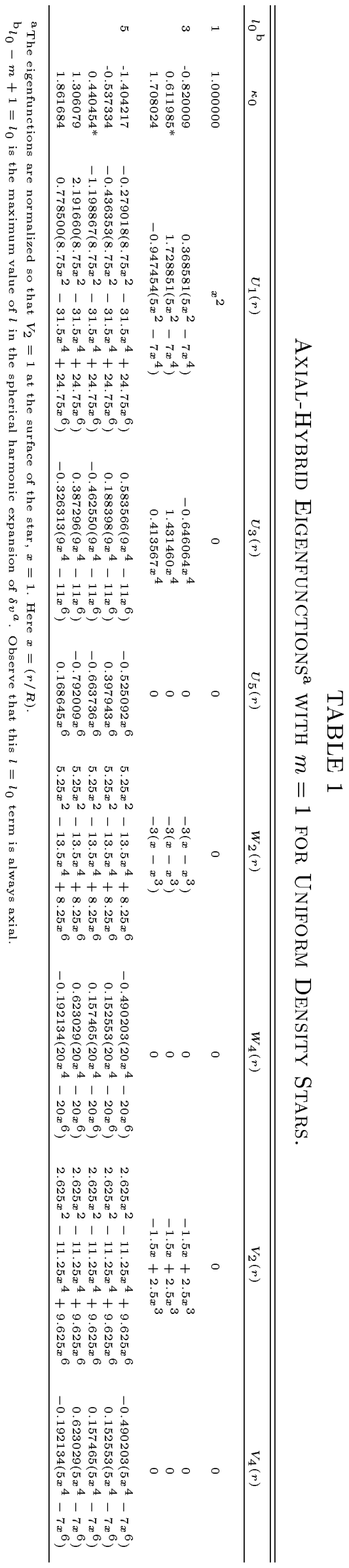}}
\includegraphics{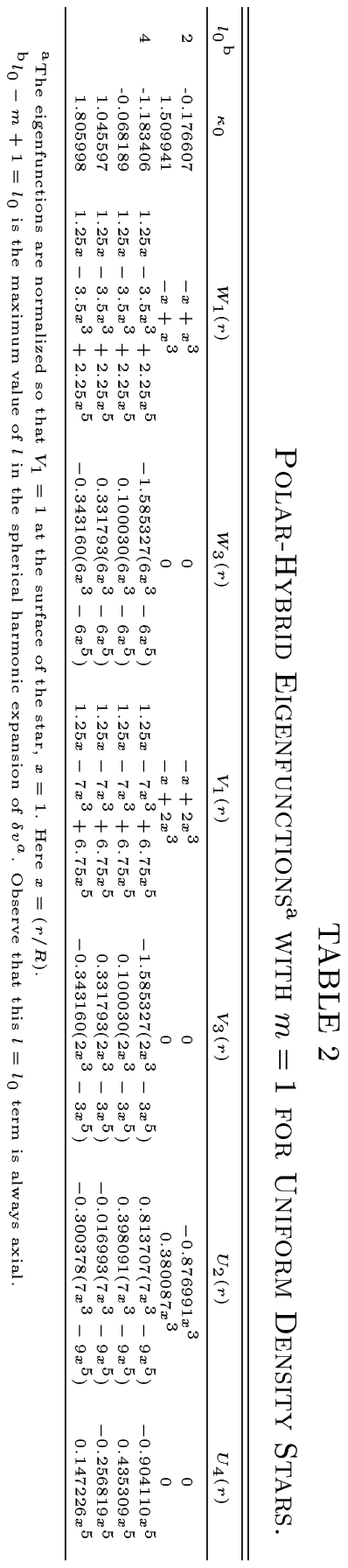}
\resizebox{\hsize}{9.5in}{\includegraphics{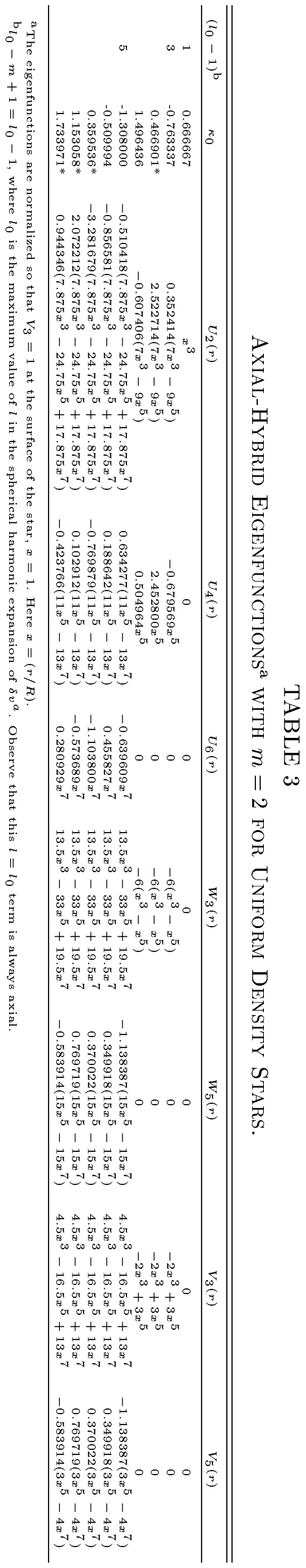}}
\includegraphics{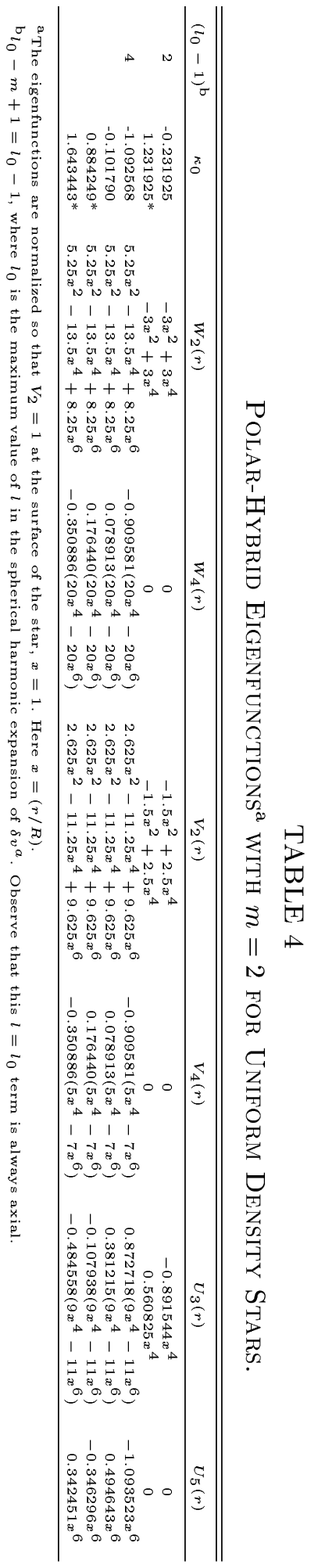}


\begin{table}
\dummytable\label{ef_m1a}
\end{table}

\begin{table}
\dummytable\label{ef_m1p}
\end{table}

\begin{table}
\dummytable\label{ef_m2a}
\end{table}

\begin{table}
\dummytable\label{ef_m2p}
\end{table}

\clearpage

\begin{deluxetable}{ccrrrrr}
\footnotesize
\tablecaption{Eigenvalues $\kappa_0\tablenotemark{a}$ \ for Uniform Density 
Stars. \label{ev_mac}}
\tablewidth{0pt}
\tablehead{
\colhead{($l_0-m+1$)\tablenotemark{b}} & \colhead{parity\tablenotemark{c}} & 
\colhead{$m=0$}   & \colhead{$m=1$}  & \colhead{$m=2$}   & \colhead{$m=3$}   & \colhead{$m=4$} 
} 
\startdata
 1\tablenotemark{d}
  &a &  0.000000 &  1.000000\phm{*}   &  0.666667*         &  0.500000*         &  0.400000*  \nl
 2&p & -0.894427 & -0.176607\phm{*}   & -0.231925\phm{*}   & -0.253197\phm{*}   & -0.261255\phm{*}   \nl
  &p &  0.894427 &  1.509941\phm{*}   &  1.231925*         &  1.053197*         &  0.927922*  \nl
 3&a & -1.309307 & -0.820009\phm{*}   & -0.763337\phm{*}   & -0.718066\phm{*}   & -0.680693\phm{*}   \nl
  &a &  0.000000 &  0.611985*         &  0.466901*         &  0.377861*         &  0.317496*  \nl
  &a &  1.309307 &  1.708024\phm{*}   &  1.496436*         &  1.340205*         &  1.220340*  \nl
 4&p & -1.530111 & -1.183406\phm{*}   & -1.092568\phm{*}   & -1.022179\phm{*}   & -0.965177\phm{*}   \nl
  &p & -0.570463 & -0.068189\phm{*}   & -0.101790\phm{*}   & -0.120347\phm{*}   & -0.131215\phm{*}   \nl
  &p &  0.570463 &  1.045597\phm{*}   &  0.884249*         &  0.773460*         &  0.691976*  \nl
  &p &  1.530111 &  1.805998\phm{*}   &  1.643443*         &  1.511923*         &  1.404416*  \nl
 5&a & -1.660448 & -1.404217\phm{*}   & -1.308000\phm{*}   & -1.230884\phm{*}   & -1.167037\phm{*}   \nl
  &a & -0.937698 & -0.537334\phm{*}   & -0.509994\phm{*}   & -0.486868\phm{*}   & -0.466934\phm{*}   \nl
  &a &  0.000000 &  0.440454*         &  0.359536*         &  0.304044*         &  0.263530*  \nl
  &a &  0.937698 &  1.306079\phm{*}   &  1.153058*         &  1.040073*         &  0.952507*  \nl
  &a &  1.660448 &  1.861684\phm{*}   &  1.733971*         &  1.623634*         &  1.529045*  \nl
\enddata
 
\tablenotetext{a}{$\kappa_0\Omega=(\sigma+m\Omega)$ is the mode
frequency in the rotating frame to lowest order in $\Omega$.  The modes whose frequencies are 
marked with a $\ast$ satisfy the condition $\sigma(\sigma+m\Omega)<0$ and are subject to a 
gravitational radiation driven instability in the absence of viscous dissipation.}
\tablenotetext{b}{For $m=0$, this is simply $l_0$. For the uniform density star, $l_0$ is the
maximum value of $l$ appearing in the spherical harmonic expansion of $\delta v^a$.}
\tablenotetext{c}{This denotes the parity class of the mode; a meaning axial-led hybrids,
and p meaning polar-led hybrids.}
\tablenotetext{d}{These are the eigenvalues of the pure $l_0=m$ r-modes. For isentropic stars
they are independent of the equation of state and have the value $\kappa_0=2/(m+1)$ (or
$\kappa_0=0$ for $m=0$) to lowest order in $\Omega$  (\cite{pp78}).}
\end{deluxetable}

\clearpage

\begin{deluxetable}{ccrrrrr}
\footnotesize
\tablecaption{Eigenvalues $\kappa_0\tablenotemark{a}$ \ for the $p=K\rho^2$ 
Polytrope. \label{ev_poly}}
\tablewidth{0pt}
\tablehead{
\colhead{($l_0-m+1$)\tablenotemark{b}} & \colhead{parity\tablenotemark{c}} & 
\colhead{$m=0$}   & \colhead{$m=1$}  & \colhead{$m=2$}   & \colhead{$m=3$}   & \colhead{$m=4$} 
} 
\startdata
 1\tablenotemark{d}
  &a &  0.000000 &  1.000000\phm{*}  &  0.666667*       &  0.500000*         &  0.400000*  \nl
 2&p & -1.028189 & -0.401371\phm{*}  & -0.556592\phm{*} & -0.631637\phm{*}   & -0.672385\phm{*}   \nl
  &p &  1.028189 &  1.412999\phm{*}  &  1.100026*       &  0.904910*         &  0.771078*  \nl
 3&a & -1.358128 & -1.032380\phm{*}  & -1.025883\phm{*} & -1.014866\phm{*}   & -1.002175\phm{*}   \nl
  &a &  0.000000 &  0.690586*        &  0.517337*       &  0.412646*         &  0.342817*  \nl
  &a &  1.358128 &  1.613725\phm{*}  &  1.357781*       &  1.176745*         &  1.041683*  \nl
 4&p & -1.542065 & -1.312267\phm{*}  & -1.272885\phm{*} & -1.238631\phm{*}   & -1.208390\phm{*}   \nl
  &p & -0.701821 & -0.178792\phm{*}  & -0.275335\phm{*} & -0.333267\phm{*}   & -0.370450\phm{*}   \nl
  &p &  0.701821 &  1.051525\phm{*}  &  0.862948*       &  0.734297*         &  0.640592*  \nl
  &p &  1.542065 &  1.726257\phm{*}  &  1.519573*       &  1.360560*         &  1.234698*  \nl
 5&a & -1.656481 & -1.483402\phm{*}  & -1.433916\phm{*} & -1.391305\phm{*}   & -1.354057\phm{*}   \nl
  &a & -1.013703 & -0.705182\phm{*}  & -0.703898\phm{*} & -0.699942\phm{*}   & -0.694498\phm{*}   \nl
  &a &  0.000000 &  0.528102*        &  0.421678*       &  0.350192*         &  0.299055*  \nl
  &a &  1.013703 &  1.281962\phm{*}  &  1.104402*       &  0.974192*         &  0.874124*  \nl
  &a &  1.656481 &  1.795734\phm{*}  &  1.627215*       &  1.489441*         &  1.375406*  \nl
\enddata
 
\tablenotetext{a}{$\kappa_0\Omega=(\sigma+m\Omega)$ is the mode
frequency in the rotating frame to lowest order in $\Omega$.  The modes whose frequencies 
are marked with a $\ast$
satisfy the condition $\sigma(\sigma+m\Omega)<0$ and are subject to a gravitational
radiation driven instability in the absence of viscous dissipation.}
\tablenotetext{b}{For $m=0$, this is simply $l_0$.  For the $n=1$ polytrope, $l_0$ is the largest 
value of $l$ that contributes a dominant term to the spherical harmonic expansion of $\delta v^a$.}
\tablenotetext{c}{This denotes the parity class of the mode; a meaning axial-led hybrids,
and p meaning polar-led hybrids.}
\tablenotetext{d}{These are the eigenvalues of the pure $l_0=m$ r-modes. For isentropic stars
they are independent of the equation of state and have the value $\kappa_0=2/(m+1)$ (or
$\kappa_0=0$ for $m=0$) to lowest order in $\Omega$  (\cite{pp78}).}
\end{deluxetable}

\clearpage

\begin{deluxetable}{ccccccc}
\footnotesize
\tablecaption{Dissipative timescales (in seconds) for $m=1$ axial-hybrid 
modes\tablenotemark{a} \ at $T=10^9K$ and $\Omega=\sqrt{\pi G \bar{\rho}}$.\label{times_m1a}}
\tablewidth{0pt}
\tablehead{
\colhead{$l_0$} & \colhead{$n$\tablenotemark{b}} & \colhead{$\kappa$} & 
\colhead{$\tilde \tau_B$\tablenotemark{c}} & 
\colhead{$\tilde \tau_S$} & \colhead{$\tilde \tau_3$} & \colhead{$\tilde \tau_5$} 
} 
\startdata
 3& 0 & 0.611985 &$ \cdots            $&$  7.67\times 10^{7} $&$ -9.79\times 10^{6}  $&$     \cdots          $\nl
  & 1 & 0.690586 &$ 5.86\times 10^{9} $&$  9.29\times 10^{7} $&$ -1.25\times 10^{8}  $&$ -1.22\times 10^{20} $\nl
 & & & & & & \nl
 5& 0 & 0.440454 &$ \cdots            $&$  2.04\times 10^{7} $&$  -\infty            $&$ -2.07\times 10^{13} $\nl
  & 1 & 0.528102 &$ 2.57\times 10^{9} $&$  3.87\times 10^{7} $&$ -2.17\times 10^{10} $&$ -5.75\times 10^{14} $\nl
\enddata
 
\tablenotetext{a}{We present dissipative timescales only for those modes that are unstable
to gravitational radiation reaction.  None of the $m=1$ polar-hybrid modes are unstable for
low values of $l_0$.}
\tablenotetext{b}{The polytropic index, $n$, where $p=K\rho^{1+1/n}$. The n=0 case 
represents the uniform density equilibrium star.}
\tablenotetext{c}{Dissipation due to bulk viscosity is not meaningful for uniform
density stars.}

\end{deluxetable}

\clearpage

\begin{deluxetable}{cccccccc}
\footnotesize
\tablecaption{Dissipative timescales (in seconds) for $m=2$ axial-hybrid 
modes at $T=10^9K$ and $\Omega=\sqrt{\pi G \bar{\rho}}$.\label{times_m2a}}
\tablewidth{0pt}
\tablehead{
\colhead{$(l_0-1)$} & \colhead{$n$\tablenotemark{a}} & \colhead{$\kappa$} & \colhead{$\tilde \tau_B$\tablenotemark{b}} & 
\colhead{$\tilde \tau_S$} & \colhead{$\tilde \tau_2$} & \colhead{$\tilde \tau_4$} & \colhead{$\tilde \tau_6$} 
} 
\startdata
1\tablenotemark{c}
 & 0  &0.666667 &$ \cdots            $&$ 4.46\times 10^{8} $&$ -1.56\times 10^{0}  $&$ \cdots $&$ \cdots $\nl
 & 1  &0.666667 &$ 2.0\times 10^{11} $&$ 2.52\times 10^{8} $&$ -3.26\times 10^{0}  $&$ \cdots $&$ \cdots $\nl
 & & & & & & & \nl
3& 0  &0.466901 &$ \cdots            $&$ 4.10\times 10^{7} $&$ -\infty             $&$ -3.88\times 10^{5} $&$ \cdots $\nl   
 & 1  &0.517337 &$ 6.43\times 10^{9} $&$ 6.21\times 10^{7} $&$ < -10^{18} $&$ -1.85\times 10^{6} $&$ -4.97\times 10^{15} $\nl
 & & & & & & & \nl
 & 0  &1.496436 &$ \cdots            $&$ 3.92\times 10^{7} $&$ -\infty             $&$ -5.85\times 10^{9} $&$ \cdots  $\nl  
 & 1  &1.357781 &$ 4.10\times 10^{9} $&$ 7.18\times 10^{7} $&$ < -10^{19} $&$ -1.60\times 10^{9} $&$ -4.35\times 10^{19} $\nl
 & & & & & & & \nl
5& 0  &0.359536 &$ \cdots            $&$ 1.34\times 10^{7} $&$ -\infty             $&$ -\infty     $&$ -1.28\times 10^{11} $\nl
 & 1  &0.421678 &$ 2.65\times 10^{9} $&$ 3.01\times 10^{7} $&$ < -10^{16} $&$ -2.01\times 10^{9} $&$ -1.15\times 10^{12} $\nl
 & & & & & & & \nl
 & 0  &1.153058 &$ \cdots            $&$ 1.32\times 10^{7} $&$ -\infty             $&$ -\infty     $&$ -3.11\times 10^{14} $\nl
 & 1  &1.104402 &$ 2.45\times 10^{9} $&$ 3.65\times 10^{7} $&$ < -10^{12} $&$ -1.37\times 10^{11} $&$ -4.89\times 10^{14} $\nl
 & & & & & & & \nl
 & 0  &1.733971 &$ \cdots            $&$ 1.31\times 10^{7} $&$ -\infty             $&$ -\infty     $&$ -1.92\times 10^{21} $\nl
 & 1  &1.627215 &$ 5.32\times 10^{9} $&$ 3.44\times 10^{7} $&$ < -10^{19} $&$ -2.30\times 10^{15} $&$ -8.33\times 10^{19} $\nl
\enddata
 
\tablenotetext{a}{The polytropic index, $n$, where $p=K\rho^{1+1/n}$. The n=0 case 
represents the uniform density equilibrium star.}
\tablenotetext{b}{Dissipation due to bulk viscosity is not meaningful for uniform
density stars.}
\tablenotetext{c}{This is the $l_0=m=2$ r-mode already studied by Lindblom et al. (1998), Owen et al. (1998),
Andersson et al. (1998), Kokkotas and Stergioulas (1998) and Lindblom et al. (1999).  The value of the bulk 
viscosity timescale for this mode is taken from Lindblom et al. (1999) who calculate it self-consistently 
using an order $\Omega^2$ calculation.}
\end{deluxetable}

\clearpage

\begin{deluxetable}{ccccccc}
\footnotesize
\tablecaption{Dissipative timescales (in seconds) for $m=2$ polar-hybrid 
modes at $T=10^9K$ and $\Omega=\sqrt{\pi G \bar{\rho}}$.\label{times_m2p}}
\tablewidth{0pt}
\tablehead{
\colhead{$(l_0-1)$} & \colhead{$n$\tablenotemark{a}} & \colhead{$\kappa$} & \colhead{$\tilde \tau_B$\tablenotemark{b}} & 
\colhead{$\tilde \tau_S$} & \colhead{$\tilde \tau_3$} & \colhead{$\tilde \tau_5$}
} 
\startdata
2& 0  &1.231925 &$   \cdots          $&$ 9.03\times 10^{7} $&$ -4.77\times 10^{4}  $&$ \cdots              $\nl
 & 1  &1.100026 &$ 3.32\times 10^{9} $&$ 1.24\times 10^{8} $&$ -3.37\times 10^{4} $&$ -3.13\times 10^{14} $\nl
 & & & & & & \nl
4& 0  &0.884249 &$   \cdots          $&$ 2.17\times 10^{7} $&$ -\infty             $&$ -5.64\times 10^{9}  $\nl  
 & 1  &0.862948 &$ 1.93\times 10^{9} $&$ 4.94\times 10^{7} $&$ -1.10\times 10^{7} $&$ -1.45\times 10^{10} $\nl
 & & & & & & \nl
 & 0  &1.643443 &$   \cdots          $&$ 2.13\times 10^{7} $&$ -\infty             $&$ -2.12\times 10^{15} $\nl  
 & 1  &1.519573 &$ 4.79\times 10^{9} $&$ 4.77\times 10^{7} $&$ -1.92\times 10^{11} $&$ -2.31\times 10^{14} $\nl
\enddata
 
\tablenotetext{a}{The polytropic index, $n$, where $p=K\rho^{1+1/n}$. The n=0 case 
represents the uniform density equilibrium star.}
\tablenotetext{b}{Dissipation due to bulk viscosity is not meaningful for uniform
density stars.}
\end{deluxetable}


\clearpage
\begingroup\makeatletter\ifx\SetFigFont\undefined%
\gdef\SetFigFont#1#2#3{%
\reset@font\fontsize{#1}{#2pt}%
\fontfamily{#3}
\selectfont}%
\fi\endgroup%
\begin{picture}(0,0)%
\includegraphics{fig1.pstex}
\end{picture}%
\setlength{\unitlength}{0.24pt}%
\begingroup\makeatletter\ifx\SetFigFont\undefined%
\gdef\SetFigFont#1#2#3{%
\reset@font\fontsize{#1}{#2pt}%
\fontfamily{#3}
\selectfont}%
\fi\endgroup%
\begin{picture}(1416,1912)(249,260)
\put(707,1035){\makebox(0,0)[lb]{\smash{\SetFigFont{15.0}{24.0}{rm} $U_2(r)$}}}
\put(707,961){\makebox(0,0)[lb]{\smash{\SetFigFont{15.0}{24.0}{rm} $W_2(r)$}}}
\put(707,887){\makebox(0,0)[lb]{\smash{\SetFigFont{15.0}{24.0}{rm} $V_2(r)$}}}
\put(707,813){\makebox(0,0)[lb]{\smash{\SetFigFont{15.0}{24.0}{rm} $U_4(r)$}}}
\end{picture}

\begin{center} Figure 1 \end{center}
\clearpage
\begingroup\makeatletter\ifx\SetFigFont\undefined%
\gdef\SetFigFont#1#2#3{%
\reset@font\fontsize{#1}{#2pt}%
\fontfamily{#3}
\selectfont}%
\fi\endgroup%
\begin{picture}(0,0)%
\includegraphics{fig2.pstex}
\end{picture}%
\setlength{\unitlength}{0.24pt}%
\begingroup\makeatletter\ifx\SetFigFont\undefined%
\gdef\SetFigFont#1#2#3{%
\reset@font\fontsize{#1}{#2pt}%
\fontfamily{#3}
\selectfont}%
\fi\endgroup%
\begin{picture}(1416,1912)(249,260)
\put(707,1275){\makebox(0,0)[lb]{\smash{\SetFigFont{15.0}{24.0}{rm} $U_2(r)$}}}
\put(707,1201){\makebox(0,0)[lb]{\smash{\SetFigFont{15.0}{24.0}{rm} $W_2(r)$}}}
\put(707,1127){\makebox(0,0)[lb]{\smash{\SetFigFont{15.0}{24.0}{rm} $V_2(r)$}}}
\put(707,1053){\makebox(0,0)[lb]{\smash{\SetFigFont{15.0}{24.0}{rm} $U_4(r)$}}}
\put(707,979){\makebox(0,0)[lb]{\smash{\SetFigFont{15.0}{24.0}{rm} $W_5(r)$}}}
\put(707,905){\makebox(0,0)[lb]{\smash{\SetFigFont{15.0}{24.0}{rm} $V_5(r)$}}}
\put(707,831){\makebox(0,0)[lb]{\smash{\SetFigFont{15.0}{24.0}{rm} $U_6(r)$}}}
\end{picture}

\begin{center} Figure 2 \end{center}
\clearpage
\begingroup\makeatletter\ifx\SetFigFont\undefined%
\gdef\SetFigFont#1#2#3{%
\reset@font\fontsize{#1}{#2pt}%
\fontfamily{#3}
\selectfont}%
\fi\endgroup%
\begin{picture}(0,0)%
\includegraphics{fig3.pstex}
\end{picture}%
\setlength{\unitlength}{0.24pt}%
\begingroup\makeatletter\ifx\SetFigFont\undefined%
\gdef\SetFigFont#1#2#3{%
\reset@font\fontsize{#1}{#2pt}%
\fontfamily{#3}
\selectfont}%
\fi\endgroup%
\begin{picture}(1574,1911)(195,260)
\put(615,1874){\makebox(0,0)[lb]{\smash{\SetFigFont{15.0}{24.0}{rm} $W_1(r)$ for the uniform density star}}}
\put(615,1800){\makebox(0,0)[lb]{\smash{\SetFigFont{15.0}{24.0}{rm} $W_1(r)$ for the $n=1$ polytrope}}}
\put(615,1726){\makebox(0,0)[lb]{\smash{\SetFigFont{15.0}{24.0}{rm} $W_3(r)$ for the $n=1$ polytrope}}}
\put(615,1652){\makebox(0,0)[lb]{\smash{\SetFigFont{15.0}{24.0}{rm} $W_5(r)$ for the $n=1$ polytrope}}}
\end{picture}

\begin{center} Figure 3 \end{center}
\clearpage
\begingroup\makeatletter\ifx\SetFigFont\undefined%
\gdef\SetFigFont#1#2#3{%
\reset@font\fontsize{#1}{#2pt}%
\fontfamily{#3}
\selectfont}%
\fi\endgroup%
\begin{picture}(0,0)%
\includegraphics{fig4.pstex}
\end{picture}%
\setlength{\unitlength}{0.24pt}%
\begingroup\makeatletter\ifx\SetFigFont\undefined%
\gdef\SetFigFont#1#2#3{%
\reset@font\fontsize{#1}{#2pt}%
\fontfamily{#3}
\selectfont}%
\fi\endgroup%
\begin{picture}(1564,1912)(195,259)
\put(615,1875){\makebox(0,0)[lb]{\smash{\SetFigFont{15.0}{24.0}{rm} $V_1(r)$ for the uniform density star}}}
\put(615,1801){\makebox(0,0)[lb]{\smash{\SetFigFont{15.0}{24.0}{rm} $V_1(r)$ for the $n=1$ polytrope}}}
\put(615,1727){\makebox(0,0)[lb]{\smash{\SetFigFont{15.0}{24.0}{rm} $V_3(r)$ for the $n=1$ polytrope}}}
\put(615,1653){\makebox(0,0)[lb]{\smash{\SetFigFont{15.0}{24.0}{rm} $V_5(r)$ for the $n=1$ polytrope}}}
\end{picture}

\begin{center} Figure 4 \end{center}
\clearpage
\begingroup\makeatletter\ifx\SetFigFont\undefined%
\gdef\SetFigFont#1#2#3{%
\reset@font\fontsize{#1}{#2pt}%
\fontfamily{#3}
\selectfont}%
\fi\endgroup%
\begin{picture}(0,0)%
\includegraphics{fig5.pstex}
\end{picture}%
\setlength{\unitlength}{0.24pt}%
\begingroup\makeatletter\ifx\SetFigFont\undefined%
\gdef\SetFigFont#1#2#3{%
\reset@font\fontsize{#1}{#2pt}%
\fontfamily{#3}
\selectfont}%
\fi\endgroup%
\begin{picture}(1570,1912)(195,260)
\put(615,1875){\makebox(0,0)[lb]{\smash{\SetFigFont{15.0}{24.0}{rm} $U_2(r)$ for the uniform density star}}}
\put(615,1801){\makebox(0,0)[lb]{\smash{\SetFigFont{15.0}{24.0}{rm} $U_2(r)$ for the $n=1$ polytrope}}}
\put(615,1727){\makebox(0,0)[lb]{\smash{\SetFigFont{15.0}{24.0}{rm} $U_4(r)$ for the $n=1$ polytrope}}}
\put(615,1653){\makebox(0,0)[lb]{\smash{\SetFigFont{15.0}{24.0}{rm} $U_6(r)$ for the $n=1$ polytrope}}}
\end{picture}

\begin{center} Figure 5 \end{center}
\clearpage
\begingroup\makeatletter\ifx\SetFigFont\undefined%
\gdef\SetFigFont#1#2#3{%
\reset@font\fontsize{#1}{#2pt}%
\fontfamily{#3}
\selectfont}%
\fi\endgroup%
\begin{picture}(0,0)%
\includegraphics{fig6.pstex}
\end{picture}%
\setlength{\unitlength}{0.24pt}%
\begingroup\makeatletter\ifx\SetFigFont\undefined%
\gdef\SetFigFont#1#2#3{%
\reset@font\fontsize{#1}{#2pt}%
\fontfamily{#3}
\selectfont}%
\fi\endgroup%
\begin{picture}(1551,1875)(214,260)
\put(615,1035){\makebox(0,0)[lb]{\smash{\SetFigFont{15.0}{24.0}{rm} $U_2(r)$ for the uniform density star}}}
\put(615,961){\makebox(0,0)[lb]{\smash{\SetFigFont{15.0}{24.0}{rm} $U_2(r)$ for the $n=1$ polytrope}}}
\put(615,887){\makebox(0,0)[lb]{\smash{\SetFigFont{15.0}{24.0}{rm} $U_4(r)$ for the uniform density star}}}
\put(615,813){\makebox(0,0)[lb]{\smash{\SetFigFont{15.0}{24.0}{rm} $U_4(r)$ for the $n=1$ polytrope}}}
\put(615,739){\makebox(0,0)[lb]{\smash{\SetFigFont{15.0}{24.0}{rm} $U_6(r)$ for the $n=1$ polytrope}}}
\end{picture}

\begin{center} Figure 6 \end{center}
\clearpage
\begingroup\makeatletter\ifx\SetFigFont\undefined%
\gdef\SetFigFont#1#2#3{%
\reset@font\fontsize{#1}{#2pt}%
\fontfamily{#3}
\selectfont}%
\fi\endgroup%
\begin{picture}(0,0)%
\includegraphics{fig7.pstex}
\end{picture}%
\setlength{\unitlength}{0.24pt}%
\begingroup\makeatletter\ifx\SetFigFont\undefined%
\gdef\SetFigFont#1#2#3{%
\reset@font\fontsize{#1}{#2pt}%
\fontfamily{#3}
\selectfont}%
\fi\endgroup%
\begin{picture}(1574,1913)(195,259)
\put(615,1874){\makebox(0,0)[lb]{\smash{\SetFigFont{15.0}{24.0}{rm} $W_3(r)$ for the uniform density star}}}
\put(615,1800){\makebox(0,0)[lb]{\smash{\SetFigFont{15.0}{24.0}{rm} $W_3(r)$ for the $n=1$ polytrope}}}
\put(615,1726){\makebox(0,0)[lb]{\smash{\SetFigFont{15.0}{24.0}{rm} $W_5(r)$ for the $n=1$ polytrope}}}
\put(615,1652){\makebox(0,0)[lb]{\smash{\SetFigFont{15.0}{24.0}{rm} $W_7(r)$ for the $n=1$ polytrope}}}
\end{picture}

\begin{center} Figure 7 \end{center}
\clearpage
\begingroup\makeatletter\ifx\SetFigFont\undefined%
\gdef\SetFigFont#1#2#3{%
\reset@font\fontsize{#1}{#2pt}%
\fontfamily{#3}
\selectfont}%
\fi\endgroup%
\begin{picture}(0,0)%
\includegraphics{fig8.pstex}
\end{picture}%
\setlength{\unitlength}{0.24pt}%
\begingroup\makeatletter\ifx\SetFigFont\undefined%
\gdef\SetFigFont#1#2#3{%
\reset@font\fontsize{#1}{#2pt}%
\fontfamily{#3}
\selectfont}%
\fi\endgroup%
\begin{picture}(1564,1912)(195,259)
\put(615,1875){\makebox(0,0)[lb]{\smash{\SetFigFont{15.0}{24.0}{rm} $V_3(r)$ for the uniform density star}}}
\put(615,1801){\makebox(0,0)[lb]{\smash{\SetFigFont{15.0}{24.0}{rm} $V_3(r)$ for the $n=1$ polytrope}}}
\put(615,1727){\makebox(0,0)[lb]{\smash{\SetFigFont{15.0}{24.0}{rm} $V_5(r)$ for the $n=1$ polytrope}}}
\put(615,1653){\makebox(0,0)[lb]{\smash{\SetFigFont{15.0}{24.0}{rm} $V_7(r)$ for the $n=1$ polytrope}}}
\end{picture}

\begin{center} Figure 8 \end{center}
\clearpage
\begingroup\makeatletter\ifx\SetFigFont\undefined%
\gdef\SetFigFont#1#2#3{%
\reset@font\fontsize{#1}{#2pt}%
\fontfamily{#3}
\selectfont}%
\fi\endgroup%
\begin{picture}(0,0)%
\includegraphics{fig9.pstex}
\end{picture}%
\setlength{\unitlength}{0.24pt}%
\begingroup\makeatletter\ifx\SetFigFont\undefined%
\gdef\SetFigFont#1#2#3{%
\reset@font\fontsize{#1}{#2pt}%
\fontfamily{#3}
\selectfont}%
\fi\endgroup%
\begin{picture}(1516,1912)(249,259)
\put(615,1155){\makebox(0,0)[lb]{\smash{\SetFigFont{15.0}{24.0}{rm} $U_2(r)$ for the uniform density star}}}
\put(615,1081){\makebox(0,0)[lb]{\smash{\SetFigFont{15.0}{24.0}{rm} $U_2(r)$ for the $n=1$ polytrope}}}
\put(615,1007){\makebox(0,0)[lb]{\smash{\SetFigFont{15.0}{24.0}{rm} $U_4(r)$ for the uniform density star}}}
\put(615,933){\makebox(0,0)[lb]{\smash{\SetFigFont{15.0}{24.0}{rm} $U_4(r)$ for the $n=1$ polytrope}}}
\put(615,859){\makebox(0,0)[lb]{\smash{\SetFigFont{15.0}{24.0}{rm} $U_6(r)$ for the uniform density star}}}
\put(615,785){\makebox(0,0)[lb]{\smash{\SetFigFont{15.0}{24.0}{rm} $U_6(r)$ for $n=1$ polytrope}}}
\put(615,711){\makebox(0,0)[lb]{\smash{\SetFigFont{15.0}{24.0}{rm} $U_8(r)$ for the $n=1$ polytrope}}}
\end{picture}

\begin{center} Figure 9 \end{center}
\clearpage
\begingroup\makeatletter\ifx\SetFigFont\undefined%
\gdef\SetFigFont#1#2#3{%
\reset@font\fontsize{#1}{#2pt}%
\fontfamily{#3}
\selectfont}%
\fi\endgroup%
\begin{picture}(0,0)%
\includegraphics{fig10.pstex}
\end{picture}%
\setlength{\unitlength}{0.24pt}%
\begingroup\makeatletter\ifx\SetFigFont\undefined%
\gdef\SetFigFont#1#2#3{%
\reset@font\fontsize{#1}{#2pt}%
\fontfamily{#3}
\selectfont}%
\fi\endgroup%
\begin{picture}(1574,1911)(195,260)
\put(615,914){\makebox(0,0)[lb]{\smash{\SetFigFont{15.0}{24.0}{rm} $W_3(r)$ for the uniform density star}}}
\put(615,840){\makebox(0,0)[lb]{\smash{\SetFigFont{15.0}{24.0}{rm} $W_3(r)$ for the $n=1$ polytrope}}}
\put(615,766){\makebox(0,0)[lb]{\smash{\SetFigFont{15.0}{24.0}{rm} $W_5(r)$ for the uniform density star}}}
\put(615,692){\makebox(0,0)[lb]{\smash{\SetFigFont{15.0}{24.0}{rm} $W_5(r)$ for the $n=1$ polytrope}}}
\put(615,618){\makebox(0,0)[lb]{\smash{\SetFigFont{15.0}{24.0}{rm} $W_7(r)$ for the $n=1$ polytrope}}}
\end{picture}

\begin{center} Figure 10 \end{center}
\clearpage
\begingroup\makeatletter\ifx\SetFigFont\undefined%
\gdef\SetFigFont#1#2#3{%
\reset@font\fontsize{#1}{#2pt}%
\fontfamily{#3}
\selectfont}%
\fi\endgroup%
\begin{picture}(0,0)%
\includegraphics{fig11.pstex}
\end{picture}%
\setlength{\unitlength}{0.24pt}%
\begingroup\makeatletter\ifx\SetFigFont\undefined%
\gdef\SetFigFont#1#2#3{%
\reset@font\fontsize{#1}{#2pt}%
\fontfamily{#3}
\selectfont}%
\fi\endgroup%
\begin{picture}(1592,1912)(195,259)
\put(615,1875){\makebox(0,0)[lb]{\smash{\SetFigFont{15.0}{24.0}{rm} $V_3(r)$ for the uniform density star}}}
\put(615,1801){\makebox(0,0)[lb]{\smash{\SetFigFont{15.0}{24.0}{rm} $V_3(r)$ for the $n=1$ polytrope}}}
\put(615,1727){\makebox(0,0)[lb]{\smash{\SetFigFont{15.0}{24.0}{rm} $V_5(r)$ for the uniform  density star}}}
\put(615,1653){\makebox(0,0)[lb]{\smash{\SetFigFont{15.0}{24.0}{rm} $V_5(r)$ for the $n=1$ polytrope}}}
\put(615,1579){\makebox(0,0)[lb]{\smash{\SetFigFont{15.0}{24.0}{rm} $V_7(r)$ for the $n=1$ polytrope}}}
\end{picture}

\begin{center} Figure 11 \end{center}

\end{document}